\documentclass[10pt]{book}%%%% latex 2e
\usepackage{sussp_TnF} %%%% change directory to suit, eg '../sussp'
\usepackage{graphicx}     %%%% most useful package
\usepackage{makeidx}
\usepackage{amssymb}
\usepackage{rotating}  % Remember to copy this to main file!
\makeindex
\newcommand{\new}{\newcommand}
\new{\pt}        {p_T}
\new{\Et}        {E_T}
\new{\MET}    {/\!\!\!\!E_T}

\begin{document}

\headings{LHC Detectors and Early Physics} % actual title
{LHC Detectors and Early Physics}%  possibly abbreviated title for the running head
{G\"unther Dissertori}
{ETH Zurich, Switzerland} 

%%%%%%%%%%%%%%%%%%%%%%%%%%%%%%%%%%%%%%
% intro
%%%%%%%%%%%%%%%%%%%%%%%%%%%%%%%%%%%%%%

\section{Introduction}
 \label{sec:dissertori_intro}

At the time of writing these proceedings\footnote{Lecture given at the
65th Scottish Universities Summer School in Physics: LHC Physics (16 August to 29 August 2009), St.\ Andrews} (January 2010), 
we have just witnessed the successful re-start (after the sudden stop of operations in September 2008) 
of the world's most powerful particle accelerator
 ever built. CERN's Large Hadron Collider (LHC) has provided
 first collisions at injection energy (450 GeV per beam) and at the world's record
 energy of 1.18 TeV/beam. Next it is expected that the centre-of-mass energy
 will be raised to 7 TeV during the year 2010.

 Thanks to the unprecedented energies and luminosities, it
 will give particle physicists the possibility to explore the TeV energy
 range for the first time and hopefully discover new phenomena, which go beyond the
 so successful Standard Model (SM). Among the most prominent new physics scenarios 
 are the appearance of one (or several) Higgs bosons, of supersymmetric particles
 and of signatures for the existence of extra spatial dimensions.

In this review I will try to sketch the basic criteria and boundary conditions which have
guided the design of the LHC detectors. The discussion will concentrate on the so-called
general-purpose experiments, ATLAS and CMS. After an overview of the detector's characteristics
and performance, I will elaborate on the expected measurements of hard processes,
with emphasis on jet and vector boson production, i.e., tests of Quantum Chromodynamics (QCD)
and Electroweak Physics. 

There exist many excellent reviews on the topics which are addressed here.
Without making an attempt to be comprehensive, I would like to mention the various
articles which can be found in a recent book edited by Kane and Pierce \cite{Kane:2008zz}. Extensive
material has been compiled by the ATLAS and CMS collaborations 
 in their Technical Design Reports on the Physics performance \cite{Aad:2009wy,Ball:2007zza}, 
 and detailed descriptions of the detectors are given in \cite{:2008zzm,:2008zzk}. 
 Finally, an excellent overview of the detector's performance and their comparison can be found 
 in \cite{Froidevaux:2006rg}.

%%%%%%%%%%%%%%%%%%%%%%%%%%%%%%%%%%%%%%
% How to design your detector
%%%%%%%%%%%%%%%%%%%%%%%%%%%%%%%%%%%%%%

\section{Design Criteria for the Detectors}
 \label{sec:dissertori_Howto}
 
 When designing large, \textit{general-purpose} detector systems such as ATLAS
 and CMS, which have to operate in collider mode at an accelerator such as the LHC,
 obviously many aspects, constraints, boundary conditions etc.\ have to be taken 
 into account. First of all, the expected Physics and the accelerator's parameters 
 drive the design in a significant way. The overall layout of the experiment is strongly
 determined by the choice of the magnet system(s). Finally, the tracking, calorimeter,
 muon and data-acquisition systems are built with the goal of optimal performance 
 under the imposed boundary conditions and with the available technologies. 
 In the following I will address these aspects and show how they have led to the experiments
 as they are built today. Here the aim is not  to go into the very detail of the (technical)
 choices of ATLAS and CMS, but rather to discuss, also via some simple calculations,
 the basic parameters and their role in the development of the detectors.
 
 Before entering the discussion, it is necessary to define
 the most relevant and often used kinematic variables and relations. The transverse momentum
 $\pt$ is defined as the component of a particle's (or jet's) momentum $\vec{p}$ perpendicular to the
 beam line, i.e., $\pt = p\,\sin\theta$, where $p=|\vec{p}|$ and $\theta$ is the
 angle w.r.t.\ the beam line. When talking about energy ($E$) deposits in calorimeters (or jets built
 out of them), the transverse energy is introduced, $\Et = E\,\sin\theta$. 
 If the energy deposits are defined as vectors (by using their directions
 w.r.t.\ the interaction point), their negative vector sum gives the Missing Transverse Energy 
 (MET, $\MET$). At hadron colliders the \textit{rapidity} $y = 0.5 \ln\left[(E+p_L)/(E-p_L)\right]$
 turns out to be a well-suited kinematic variable,
 because differences in rapidity are invariant under boosts along the beam direction. Here
 $p_L$ denotes the momentum component along the beam line. For massless 
 particles the rapidity is equal to the \textit{pseudo-rapidity} $\eta = - \ln(\tan(\theta/2))$ 
 \footnote{Thus, polar angles of $\theta=1,10,90^\circ$ correspond to $\eta = 5,2.4,0$.}. Indeed,
 the detector elements at hadron colliders are typically segmented in (pseudo-)rapidity
 intervals. Finally, the azimuthal angle around the beam direction is usually denoted
 as $\phi$ or $\Phi$. 
  
 \begin{figure}[htbp]
\centering
\includegraphics[width=8cm]{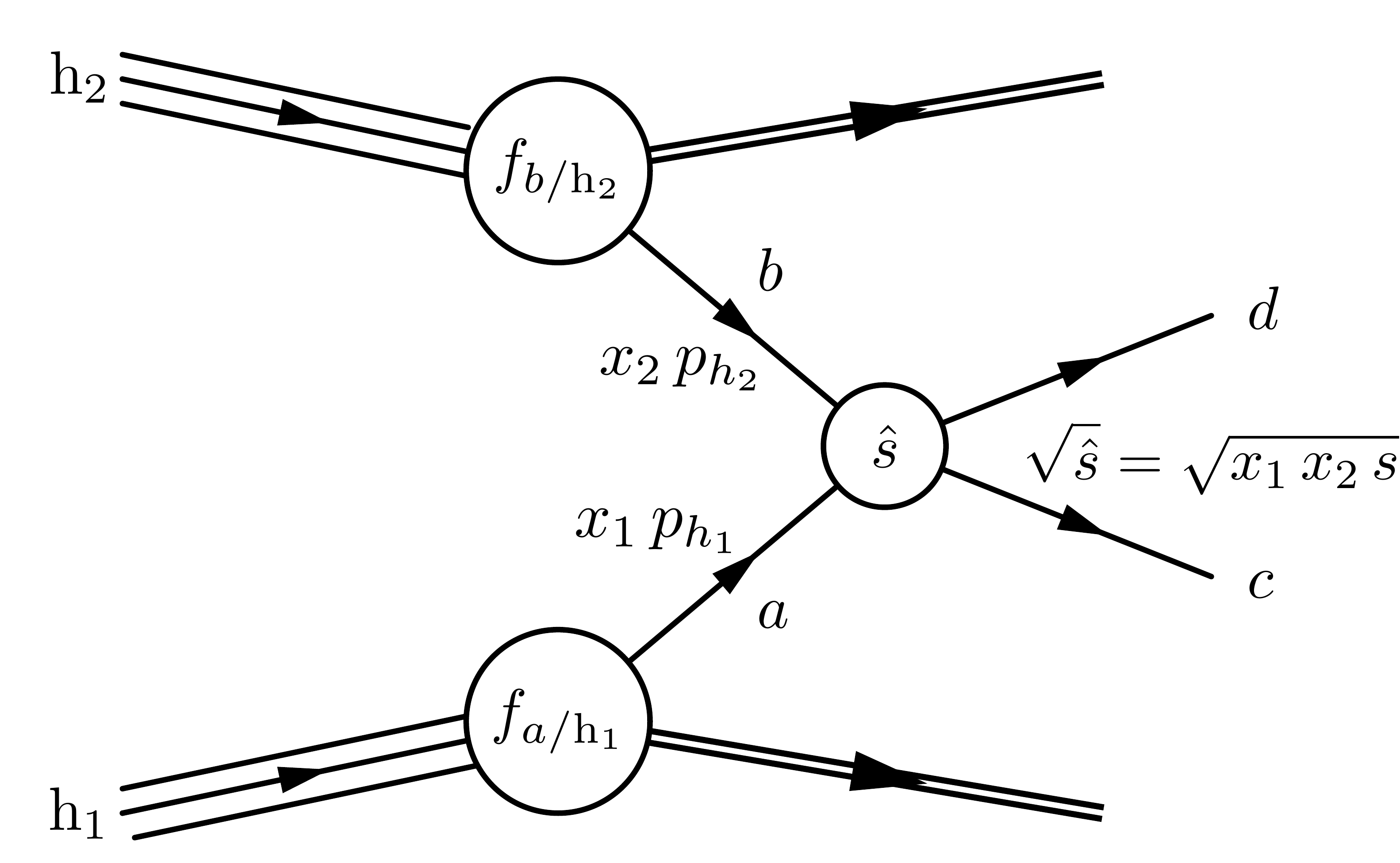}
\caption{Basic Feynman diagram for the description of inelastic hadron--hadron 
              scattering.}
\label{fig:dissertori_BasicFeynGraph}
\end{figure}

With these definitions at hand, we can now look at the basic
 kinematic relations relevant for the discussion of inelastic proton--proton scattering.
Figure \ref{fig:dissertori_BasicFeynGraph} shows the basic Feynman diagram for
 the description of inelastic hadron--hadron scattering. Two partons (quarks or gluons),
 which carry a fraction $x_{1,2}$ of the incoming hadron's momenta $p_{1,2}$, interact
 at a momentum transfer (or invariant mass) scale of $\hat s = x_1 x_2 s$,
 where $s$ is the squared centre-of-mass energy of the incoming hadrons. In the case of a heavy
 particle (or resonance) of mass $M$ 
 being produced in the interaction of the two partons, we need $\hat s = M^2$. The energy and
longitudinal momentum of the newly produced state are given by $E = (x_1+x_2)\sqrt{s}/2$
and $p_L = (x_1- x_2)\sqrt{s}/2$ (we neglect parton masses). By inserting this into the
definition for the rapidity we obtain the relations
\begin{equation}
 \label{eq:dissertori_kin}
  e^y = \sqrt{\frac{x_1}{x_2}}  \quad , \quad x_{1,2} = \frac{M}{\sqrt{s}} e^{\pm y} \; ,
\end{equation}
where $y$ is the rapidity of the particle with mass $M$. If $M$ is very large (of the order
of the total centre-of-mass energy), both momentum fractions have to be very large (and not
too dissimilar), leading
to $e^y \rightarrow 1$, i.e., such states are produced at so-called central rapidity
($y\rightarrow 0$). On the other
hand, if the momentum fractions $x_i$ are very different, e.g.\ one of them is much smaller than the other,
then the produced final state will have a strong boost and appear at large (positive or negative)
rapidities. As a consequence, the decay products of heavy states (or particles from reactions at very large
momentum transfer) will tend to appear at 
smaller rapidities (and thus ``more centrally'', i.e., at larger polar angles) than those from the bulk of
softer interactions.

%%%%%%%%%%%%%%%%%%%%%%%%%%%%%%%%%%%%%%
% expected physics
%%%%%%%%%%%%%%%%%%%%%%%%%%%%%%%%%%%%%%
\subsection{Expected Physics}
 \label{sec:dissertori_ExpectedPhysics} 
 
\begin{figure}[htbp]
\centering
\includegraphics[width=9.5cm]{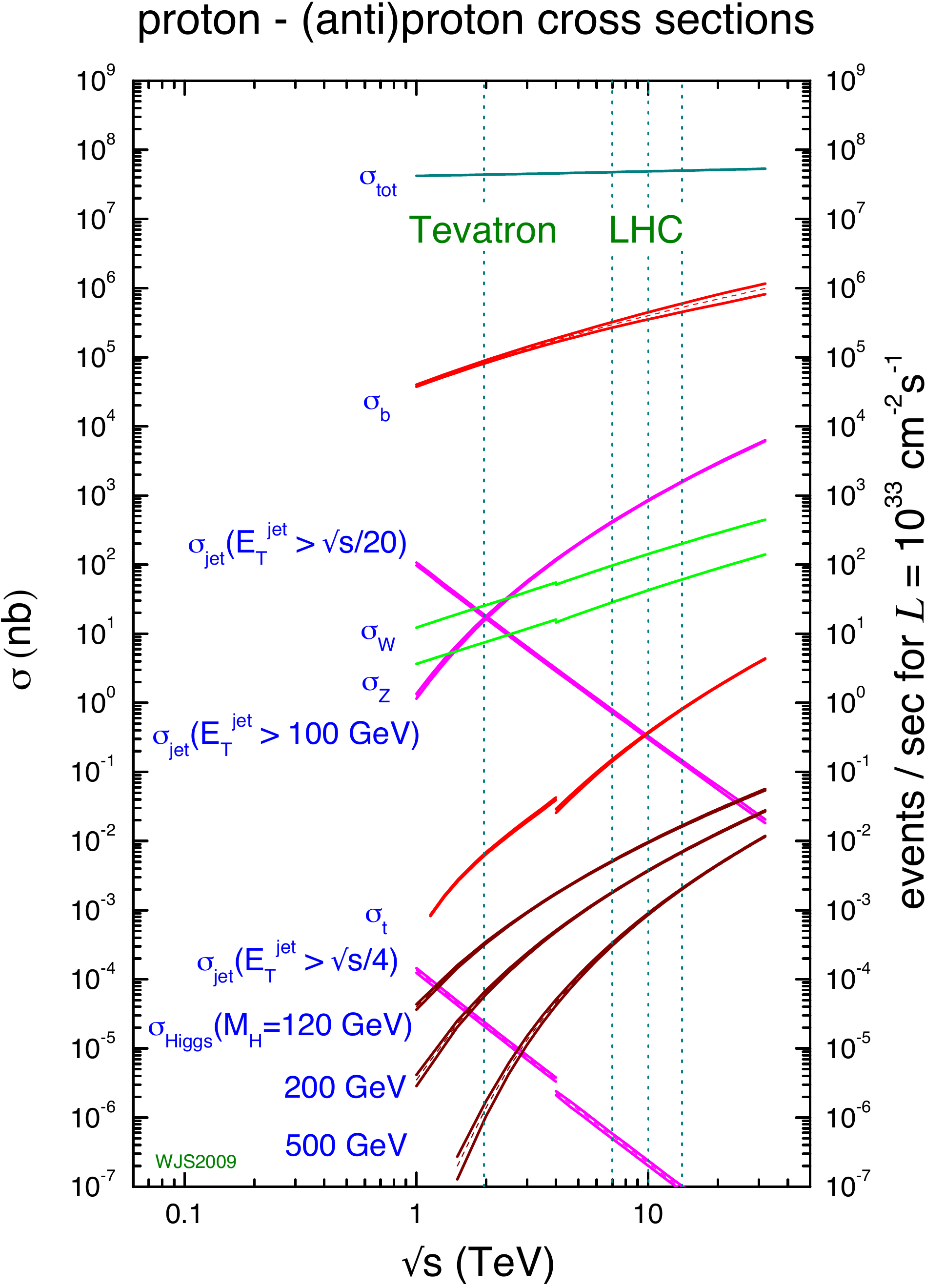}
\caption{Cross sections for various processes in (anti-)proton--proton collisions
              as a function of centre-of-mass energy, $\sqrt{s}$ (from \cite{Stirling-private}).}
\label{fig:dissertori_CrossSections}
\end{figure}

The main aspects of the Physics to be expected can be understood by studying
figure \ref{fig:dissertori_CrossSections}. There we find the cross sections for various
processes as a function of the proton--proton centre-of-mass energy. The total 
inelastic cross section, of the order of 60 - 70 mb, is completely dominated by soft
collisions at low momentum transfer. Typically this bulk of interactions is called 
\textit{minimum bias events}, since they will be registered with a minimal set of trigger conditions.
Obviously, since these are the processes with the highest cross section (probability), they
will show up first and require only a minimal selection.
Depending on the exact geometrical acceptance and the detailed definition of these
minimal trigger settings, also single- and double-diffractive events will contribute here, whereas the
measurement of purely elastic proton--proton scattering relies on dedicated detectors in the
far-forward part of the interaction region. It is worth noting that also hard-scattering
events at large momentum transfer will pass the minimum bias trigger, and thus, in principle,
 are part of the minimum bias event sample. However, 
 since their rate is so much smaller, the sample is basically populated by relatively soft collisions.  
 
 The average transverse momentum of charged
particles produced in minimum bias events at 14 TeV centre-of-mass energy
will be around 600 MeV and the most likely transverse momentum
 in the 200-300 MeV range (the exact values strongly depend
on the model parameters, which are not well known so far, or on the extrapolations
of fits to lower-energy data). From pure phase-space considerations, 
$d^3p/(2E) = \pi/2\, d\pt^2\, dy$, we expect a basically flat distribution of charged particles in rapidity,
up to the kinematic boundaries of around 5 units in rapidity where there is a sharp drop. Whereas
earlier measurements (as well as the very first LHC measurements 
at 900 and 2360 GeV centre-of-mass energy \cite{Collaboration:2009dt,CMS-dNdeta})
give slightly more than three charged particles per unit rapidity for the central plateau,
at the higher LHC energies we expect up to six charged and 2--3 neutral particles per unit rapidity,
uniformly distributed in $\phi$. These numbers allow us to make the following conclusions:
\begin{itemize}
	\item roughly integrating the charged particle densities over the whole acceptance, we obtain
	$(2\times y_\mathrm{max}) \times (6+3) \approx 90$ particles per minimum bias event, for
	   a maximum rapidity coverage of $y_\mathrm{max} =5$, and a total transverse momentum of
	     $\approx 90\times 0.6 = 54$ GeV. If several minimum bias events occur at the same
	      beam crossing, these numbers have to be multiplied accordingly. Thus it is important to have detectors
	      with excellent coverage (hermeticity) up to rapidities of around 5, in order to collect the largest
	      fraction of these particles and thus to avoid spurious measurements of $\MET$;
	\item it will be important to actually measure these numbers and their energy dependence as 
	    quickly as possible, in order to tune the Monte Carlo predictions, which later will be used to
	     model the contributions of such soft events to the energy flow in overlapping soft
	      and hard-scattering proton--proton collisions;
	\item in a strong solenoidal magnetic field, many of these soft charged particles will become curling
	     tracks (``loopers''). If we would like to avoid too many of the soft particles to reach the calorimeters,
	      thus reducing the occupancy there, we can use the simple equation 
	       $p[\mathrm{GeV}]=0.3\, R [\mathrm{m}]\, B [\mathrm{T}]$ to derive a minimal radius 
	       $L \le 2 R = p/(0.15\, B) \approx 1$ m of a cylindrical tracking detector inside a calorimeter system.
	        Here $R$ denotes the bending radius of charged particles in a magnetic field $B$. We have 
	        used $p \approx 0.6$ GeV and $B=4$ T. Stronger fields and/or larger tracking systems will 
	        keep the calorimeters cleaner from these soft-event contributions, but induce more
	        loopers, thus higher occupancy and more difficult pattern/track recognition in the tracker.
\end{itemize}

The next most-likely events to appear will be collisions containing two or more jets. Depending on
the minimal threshold in $\Et$, applied for triggering on such jets, the cross sections can grow
up to hundreds of $\mu$b. More interestingly, even the cross section for the production of
jets above several hundreds of GeV are in the tens-of-nb range and thus
 much larger than at the Tevatron. Therefore LHC will quickly extend the jet $\Et$ range into
 uncharted territory and become sensitive to high-mass di-jet resonances, new contact interactions 
 and/or quark compositeness. However, achieving such Physics goals quickly will require an
 excellent understanding of jet reconstruction. This vast topic includes the usage of modern
 jet algorithms (for a recent overview we refer to \cite{Salam:2009jx}), the data-driven determination of
 the jet energy scale, the combination of calorimeter and tracker information for obtaining an optimal
 jet energy scale and resolution, the understanding of pile-up energy and underlying event contributions, etc.
 The resulting detector requirements are an excellent (granular) calorimeter system up to
 large rapidities, combined with
 a high-performance tracking detector if the overall (charged and neutral) particle-flow reconstruction 
 is used as input to the jet finding.
 
 Once the integrated luminosities allow to be sensitive to processes with cross sections in the
 nb range, we start to explore the domain of electro-weak physics, starting with the production
 of $W$ bosons (plus anything), followed by $Z$ and then top-pair production. The latter is up to 100 times larger than
 at the Tevatron, which justifies calling the LHC a top-factory. Since the cross sections for QCD jet production
 are many orders of magnitude larger than for these electro-weak processes, it is
 obviously hopeless to look for $W$ and $Z$ decays into jets. Vector boson production is rather triggered on 
 by their leptonic decays. This explains why a major effort went into the design of detectors
 with excellent lepton (electron, muon, tau) reconstruction capabilities. Further arguments for this
 can be found by looking at a list of benchmark processes, which were identified in the early days
 of the LHC planning to be well-suited for the study of electro-weak symmetry breaking, namely
 $pp\rightarrow W^+W^-\rightarrow \mu^+\nu_\mu\mu^-\bar\nu_\mu$, 
 $pp\rightarrow Z Z\rightarrow \mu^+\mu^-\mu^+\mu^-$,
 $pp\rightarrow Z Z\rightarrow \mu^+\mu^-\nu_\mu\bar\nu_\mu$,
 $pp\rightarrow H\rightarrow\gamma\gamma$,
 $pp\rightarrow H\, \mathrm{jet}\,\mathrm{jet}$ (Vector Boson Fusion, VBF), as well as new Physics such as
 $pp\rightarrow Z'\rightarrow \mu^+\mu^-$. All these processes have cross sections (times branching
 ratios) of order $1 - 100$ fb, which immediately determines the necessary integrated luminosities for
 obtaining sensitivity to them. Again, this requires excellent lepton, photon and jet reconstruction. Since 
 here the leptons (photons) are produced in decays of heavy objects, they will be triggered on by their
 relatively large transverse momentum (typically above $20$ GeV). Furthermore, as we have seen above,
 they tend to be produced at central rapidities, thus the relevant sub-detectors, such as the
 electromagnetic calorimeter or the muon system, have to be optimized for and cover only a restricted
 (pseudo)-rapidity range, typically up to $\eta =2.5$. Altogether the above considerations lead us to the
 following additional detector requirements~:
 \begin{itemize}
\item the detectors must be capable of triggering on and identifying extremely rare events, with
    cross sections some $10^{-14}$ times the total cross section. This requires an  
     online rejection of $\sim 10^7$. Nevertheless, the overall event rates are very large and
     impose strong requirements on data handling and storage, with
     a yearly yield of $\sim 10^9$ events of a few MByte each;
 \item leptons and photons should be well triggered on and measured with high resolution. This determines
     the performance of the electromagnetic calorimetry and the muon systems. In particular, an
     efficient  electron reconstruction together with a jet fake rate\footnote{This is the probability that
     a jet is mis-identified as an electron by the reconstruction algorithms.} of about $10^{-5}$ has to be achieved, 
     in order to cope with the
     enormous jet background. Furthermore, an excellent energy resolution of the electromagnetic calorimeter
     will ease the identification of resonances over a very large background, 
     such as $H\rightarrow\gamma\gamma$;
  \item many of the electro-weak processes, such as top production, as well as new physics scenarios
   such as supersymmetric Higgs decays, involve the production of b-quarks and taus. This asks for
     silicon strip and pixel detectors, which give efficient b-tagging and tau identification. Here, jet
     rejection factors of at least $\sim 100$ are needed, for a b-tagging efficiency of $\sim 50\%$.
\end{itemize}
 
 %%%%%%%%%%%%%%%%%%%%%%%%%%%%%%%%%%%%%%
% machine parameters
%%%%%%%%%%%%%%%%%%%%%%%%%%%%%%%%%%%%%%
\subsection{The LHC Parameters}
 \label{sec:dissertori_LHCparam} 

Evidently, the LHC machine parameters play a fundamental role in the design
of the experiments. The ultimate centre-of-mass energy of 14 TeV will be reached in
several steps, currently the most likely scenario is $0.9 \rightarrow 2.36 \rightarrow 7 \rightarrow 10 \rightarrow 14$ TeV.
The maximal energy is basically fixed by the radius of the LEP tunnel and the available 
superconducting magnet technologies. During the early planning phases it became a clear
design goal that the lower energy compared to the SSC in the US (20 TeV/beam) had to be
compensated by a much higher luminosity.  Previously we have seen that processes related to the electro-weak symmetry breaking have cross sections of order $1 - 100$ fb. If we assume a canonical running time per year of $T=10^7$ sec, we will need an 
instantaneous luminosity of $\mathcal{L} = 10^{34}$ cm$^{-1}$sec$^{-1} = 10^{-5}$ fb$^{-1}$sec$^{-1}$
in order to accumulate $N=(\mathcal{L}\cdot T)\,\sigma \sim 100$ events per year for a process with
cross section $\sigma = 1$ fb.  Luminosities in this range can only be achieved by 
 large bunch intensities ($\sim 10^{11}$ protons/bunch), a large number of bunches
 or equivalently a small bunch spacing ($25$ ns) and a small beam size at the interaction regions
 ($\mathcal{O}(15-20) \mu$m). It is worth noting that a simple multiplication of instantaneous peak luminosity
and running time gives a too optimistic estimate for the integrated luminosity. Decreasing luminosities
during a fill, the time needed for the filling and acceleration cycles, 
machine commissioning and development, as well as other down-time periods can be accounted for
by a heuristic efficiency factor of roughly $0.2$.

At such high luminosities of $\mathcal{O}(10^{34}$ cm$^{-1}$sec$^{-1})$, 
due to the very large total cross section, we expect the enormous rate
of inelastic events  of about $R=\sigma_{\mathrm{inel}}\,\mathcal{L} \approx \mathcal{O}(100)\,
\mathrm{mb}\times(10^7\,\mathrm{mb}^{-1} / \mathrm{sec}) = 10^9$ events / sec.  
This allows us to calculate the number of inelastic events per bunch crossing, namely
$10^9 / \mathrm{sec} \times 25\,10^{-9}\,\mathrm{sec} = 25$ events, which will pile-up on top
of a possibly interesting high-$\pt$ scattering event. These pile-up events will be soft proton--proton
interactions, simply because it is the most likely thing to happen. Previously we have estimated
that about $90$ particles are produced in such a soft collision, with a total $\pt$ of 
order $50$ GeV. Multiplying this by the number of pile-up events, we therefore expect more than
$2000$ particles carrying a total $\pt$ of more than $1$ TeV per bunch crossing. Good coverage 
and hermeticity become an even stronger requirement. At the same time, it is clear that
dealing with these pile-up events in the detectors (and individual detector channels), as well as with
the induced radiation levels, will lead to strong boundary conditions for the experiment's design. 
Indeed, at the LHC design luminosity we expect ionising doses of $\sim 2\,10^6$ Gy / $r_\perp^2$ /year,
where $r_\perp$ is the transverse distance (in cm) to the beam. Damage can also be caused by
photons created in electromagnetic showers and by the very high neutron fluences, in particular in
the forward regions (up to $10^{17}$ n/cm$^2$ over 10 years of LHC running).

The high bunch-crossing frequency, combined with the large-sized detectors, imposes a further technical challenge, related to the timing of the trigger and readout. Interactions occur every 25 ns. During such
an interval, the produced particles travel a distance of roughly 7.5 m. This is to be compared to a 
typical LHC detector radius exceeding 10 m, and overall half-length beyond 20 m. Thus, the particles
created in one or two previous crossings have not yet left the detector when the next ones are produced at
the interaction point. Furthermore, one has to consider that the electronic signals from the detectors
travel some 5 m during a 25 ns interval, and typical cable lengths are of order 100 m. 

Altogether, these challenging conditions ask for highly granular and ra\-diation-hard detectors, combined
with fast readout (20-50 ns response time). 
High granularity helps to minimize pile-up effects in a single detector element. However,
detectors with many channels (e.g.\ about 100 million pixels, 200 000 cells in an electromagnetic calorimeter)
represent a strong cost factor.

%%%%%%%%%%%%%%%%%%%%%%%%%%%%%%%%%%%%%%
% Magnet system
%%%%%%%%%%%%%%%%%%%%%%%%%%%%%%%%%%%%%%
\subsection{The Choice of the Magnet System}
 \label{sec:dissertori_Magnet}

\begin{figure}[tbhp]
\centering
\begin{tabular}{lr}
\includegraphics[width=5cm]{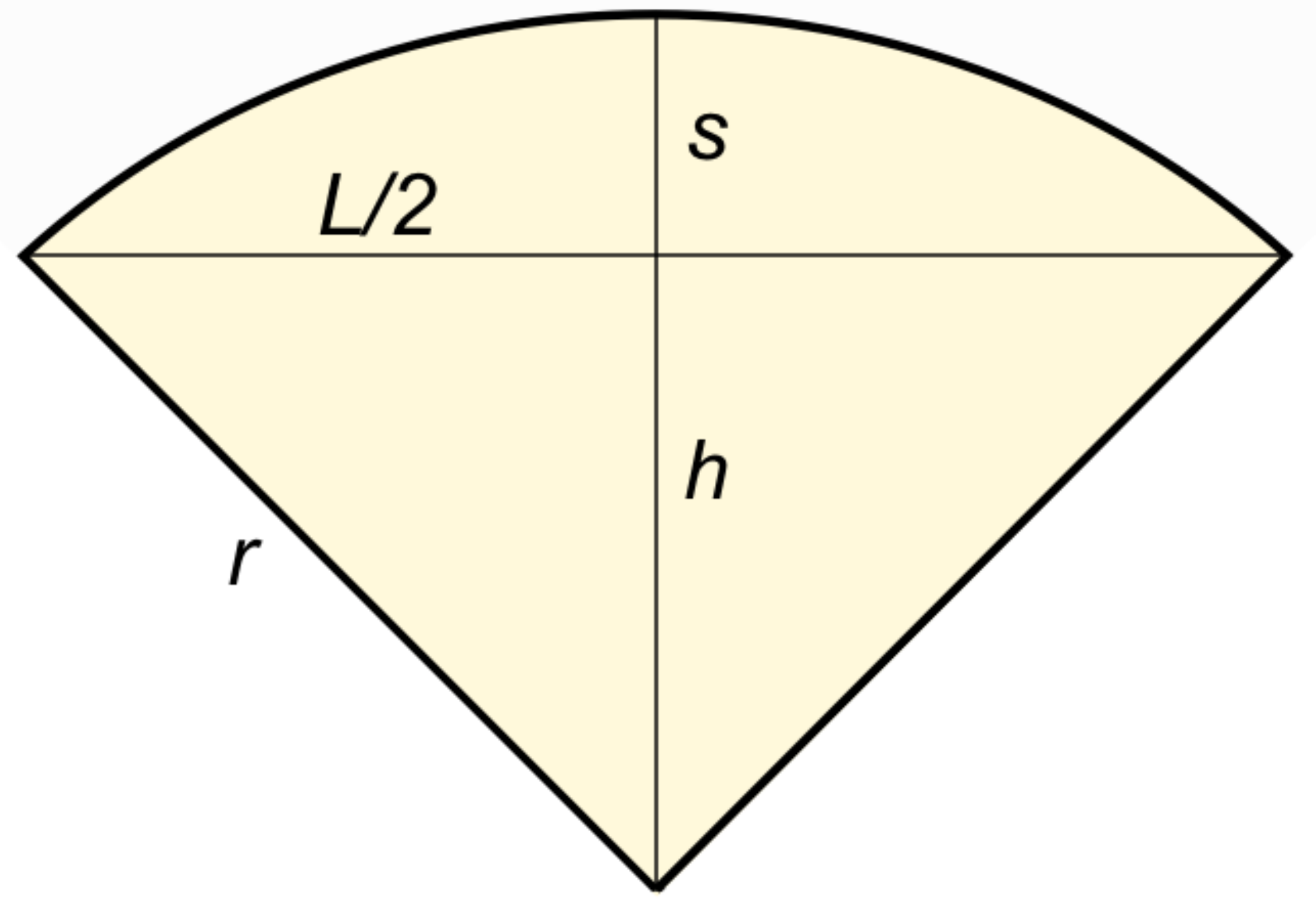}
&
\includegraphics[width=5cm]{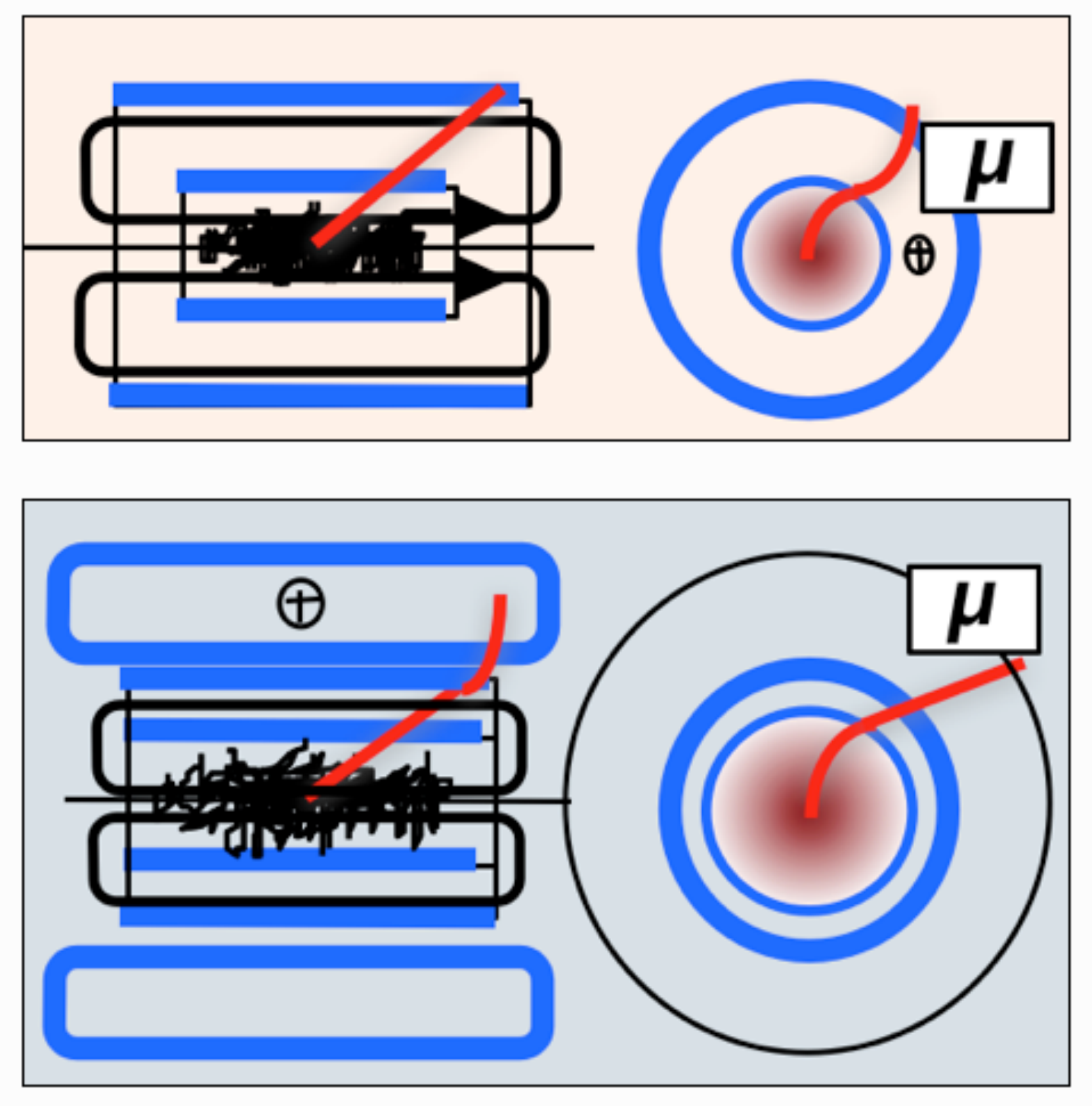} 
\end{tabular}
\caption{Left: Definition of the sagitta $s$, as used in momentum measurements based on
  the curvature of charged particle tracks in magnetic fields; Right: Two possible magnetic 
  field configurations for large particle detectors. On top a single large solenoidal field (CMS),
  at the bottom a smaller solenoid combined with a toroidal system (ATLAS). The field lines are in black,
  the coil windings are drawn in blue and the red line indicates a charged particle track.
  The black cross indicates a vector 
  field orientation perpendicular to the drawing plane  (from \cite{Cittolin}).}
\label{fig:dissertori_Magnets}
\end{figure}

The layout of the magnet system is among the most important of all design choices,
since it fixes many other parameters of the experiment. Therefore it also has to come
very early on in the development. Of course, the main purpose of the magnetic field(s)
is to bend charged particles and thus to determine their momentum and their
charge sign. Furthermore, strong fields help to keep most of the soft particles (cf.\ section
\ref{sec:dissertori_ExpectedPhysics}) within 
cylindrical regions of small radius (see figure \ref{fig:dissertori_Magnets}, right), 
thus reducing the occupancy and pile-up
effects in the calorimeters. Concerning the momentum measurement, let's recall the
most relevant formulas. Typically the momentum of a charged particle track is not determined
directly from the bending radius, but from the sagitta $s$ of a track's segment within a 
detector region of length $L$ (see figure \ref{fig:dissertori_Magnets}, left). Within the
approximation $r \gg L/2$ we find $s=L^2/(8r)$. Then the momentum $p$
and its relative uncertainty $\delta p$ are given by
\begin{equation}
  \label{eq:dissertori_Sagitta}
  p = \frac{0.3\,L^2\,B}{8\,s} \quad \Rightarrow \quad 
  \frac{\delta p}{p} = \frac{\delta s}{s} = \frac{8}{0.3}\,\frac{1}{L^2\,B}\,p\,\delta s \; ,
\end{equation}
where $B$ is the magnetic field strength (expressed in Tesla). We see that the
measurement error can be minimized by maximizing the product $L^2\,B$, i.e., it is
best to have large tracking systems (large lever arm) and strong fields. Obviously, both
parameters will drive the overall cost of the experiment, in particular $L$. It is also
worth noting that the total bending power is proportional to $\int B\, dl_\perp$, where
$l_\perp$ is the particle's path perpendicular to the magnetic field. With these
basic relations in hand, we can discuss the choices made by ATLAS and CMS
(figure \ref{fig:dissertori_Magnets}, right). A detailed comparison and technical
parameters can be found, e.g., in table 3 of Ref.\ \cite{Froidevaux:2006rg}.

  CMS has opted for a single magnet system based on a large, high-field superconducting solenoid 
  and an instrumented  iron return yoke. This has led to a simple and compact overall design
  (determining by the way also part of the experiment's name), giving excellent momentum
  resolution when combined with a powerful inner tracking system. The radius $R$ of the solenoid was
  its main cost driver, and $R\sim 3$ m turned out to be affordable and technically doable. A magnetic
  field of $4$ T was realizable, whereas $3.5$ T would still deliver good physics performance. The current
  operating field of CMS is $3.8$ T. A single solenoid has the disadvantage of limiting the 
  momentum resolution in the forward direction, at large rapidities (remember  $\int B\, dl_\perp$),
  and choosing a large winding radius implies making the solenoid also very long in order
  to cover the largest possible rapidity. The instrumented return yoke (iron) limits the momentum
   resolution at low $p$ because of multiple scattering. This also impacts the trigger rates
   and the choice of the lower trigger thresholds. This is because the steeply falling momentum
    distributions, folded with a bad resolution, lead to a large feedthrough of soft particles beyond
    the trigger thresholds and thus to saturation. This is of particular concern for the stand-alone
    muon triggering at the foreseen large Super-LHC (SLHC) rates, since the multiple scattering
    effect cannot be overcome by installing additional and/or more precise muon tracking stations. 
    A further important choice was to make the solenoid radius large enough in order to place
    the complete calorimetry (both electromagnetic and hadronic) inside the coil. Whereas this
    has the advantage of minimizing the amount of dead material in front of the calorimeter and therefore
    not compromising its intrinsic resolution, it implies that less than 2 m of radial extension
    are left for placing all absorbers, active calorimeter materials and readout, given that 
    the tracker has a radius somewhat larger than 1 m. The implications are discussed in 
    section \ref{sec:dissertori_Calo}.

    ATLAS is characterized by a two-magnet system, with a solenoid (smaller radius than in CMS)
    in the inner part and a large air-core toroid surrounding the calorimeters. In view of an
    optimal performance with muons, a toroid has the advantage of very large $L^2 B$ and good
    bending power also in the forward direction. The problems related to multiple scattering and
    its impact on muon stand-alone triggering in a high-rate environment, as discussed above, can
    be minimized by an air-core toroid, which has no return yoke. It also keeps the calorimeters
      free of field and leaves large enough space for them. However, it
    determines the extremely large overall size of ATLAS and leads to a rather complex structure
    and complex magnetic field configurations. The latter is of particular concern in view of 
    precise particle tracking, which needs excellent knowledge of the field map. Many coils
    would give a more uniform field, but obviously drive the cost very much. Indeed, the original
    proposal of 12 coils had to be reduced to 8 coils because of this. 
    Furthermore, the large magnet structures ask for very large muon
    chambers, which then have to be aligned at precisions up to $30 \mu$m, a formidable task.
    Because of the closed field lines around the calorimeters, a toroidal system has to be
    complemented by a solenoid which provides a field for inner tracking close to the interaction
    point. Since this solenoid is placed in front of the liquid-argon electromagnetic calorimeter, 
    a lot of design work went into optimizing the materials and integrating the vacuum 
    vessel for the coil with the cryostat for the barrel liquid-argon calorimeter, in order to minimize
      the dead material in front of the electromagnetic calorimeter. Also, the ATLAS solenoid is
      shorter than the one in CMS, which impacts somewhat the inner tracking performance 
      at large rapidities because of the reduced field uniformity.

%%%%%%%%%%%%%%%%%%%%%%%%%%%%%%%%%%%%%%
% Tracking
%%%%%%%%%%%%%%%%%%%%%%%%%%%%%%%%%%%%%%
\subsection{Tracking and Muon Systems}
 \label{sec:dissertori_TrackingMuon} 

The basic requirements of a tracking system are:
\begin{itemize}
\item allow for a robust and redundant pattern recognition, which
 is necessary for an efficient and precise reconstruction of all charged
  particles with momentum above $\sim 0.5$ GeV, up to pseudo-rapidities of $\sim 2.5$;
\item provide high-level triggering capabilities for electrons, taus and b-jets;
\item allow for an efficient and precise reconstruction of secondary vertices and
 impact parameters, which is of paramount importance for final states involving heavy flavours,
  in particular b-quarks;
\item complement and improve the electron reconstruction and triggering performance of the
  electromagnetic calorimeter by matching isolated tracks to calorimeter clusters;
\item provide some particle identification power, such as electron/pion separation, by
  a measurement of the specific ionization ($dE/dx$) or some other techniques, such as 
   transition radiation.
\end{itemize}

When designing a tracking system, we have to deal with a fundamental problem of 
``conflicting interests'': many tracking layers will provide many hits for a robust track reconstruction.
  However, many channels will also require a large number of supports (cables, cooling, support 
  structures etc.), which leads to a considerable amount of (dead) material in front of the calorimeters.
  This obviously jeopardizes the intrinsic resolution of the electromagnetic calorimeter, leads
   to a large fraction of photon conversions inside the tracker before reaching the calorimeter and 
   causes multiple scattering of low-momentum particles. Indeed, the issue of the so-called
   material budget of the tracking systems has led to an ``unfortunate similarity" between ATLAS
   and CMS, since in both cases the total material has increased by a factor of $\sim 2 - 2.5$ from
   their approval in 1994 to now. The material distributions reach peak levels of $\sim 1.5 - 2$ radiation lengths
   at rapidities
    around 1.5 (the transition region between barrel and endcap tracking stations is a preferred
    cable and support routing area). The consequences are that electrons lose between 25 and
    70\% of their energy by bremsstrahlung and that 20 to 60\% of the photons convert into
    $e^+e^-$ pairs before reaching the electromagnetic calorimeter. There exist algorithms for
    the recovery of energy loss by bremsstrahlung and for finding conversions, but obviously the overall reconstruction
    of electrons and photons is hampered by this.
    
 \begin{figure}[htbp]
\centering
\includegraphics[width=10cm]{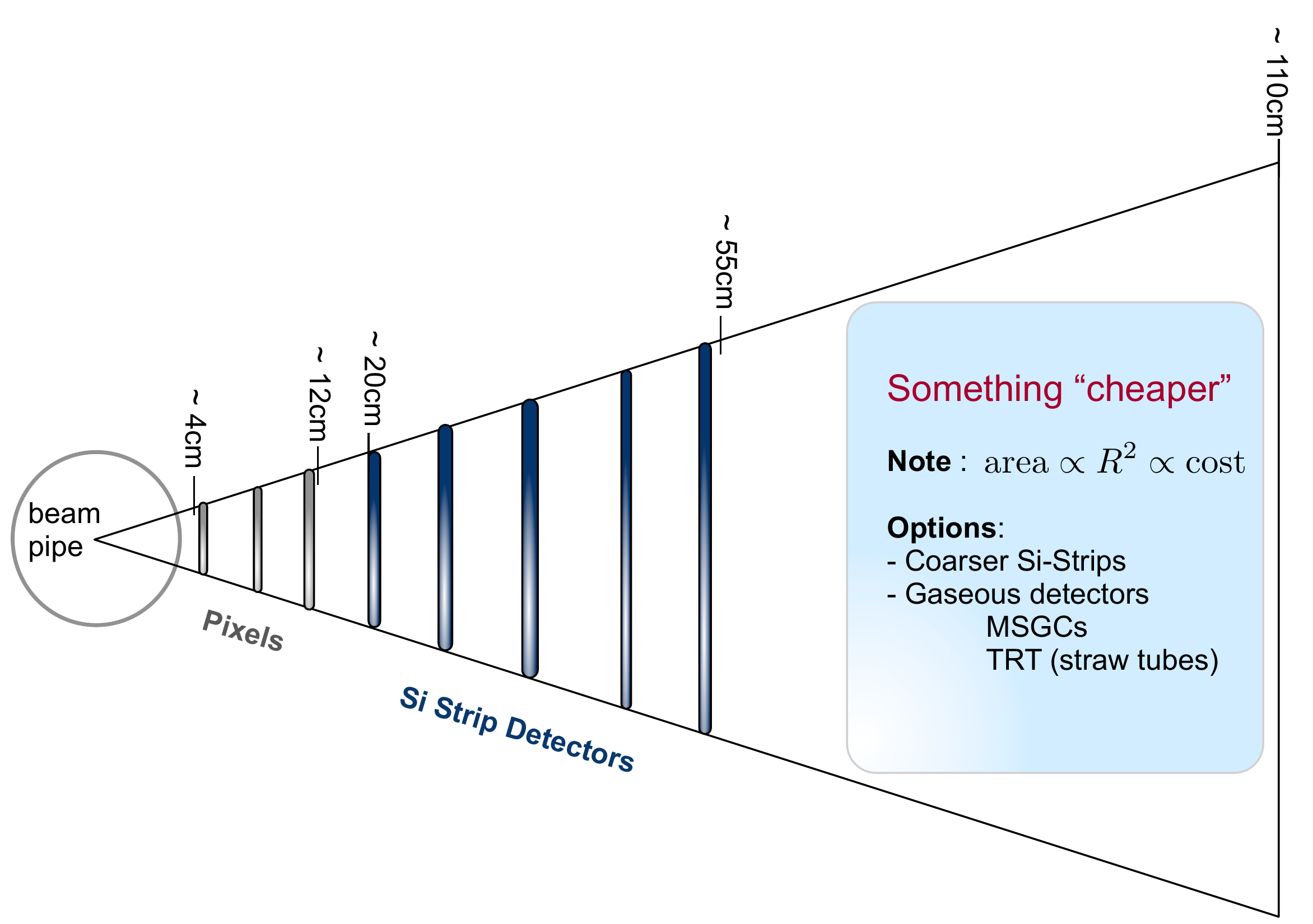} 
\caption{Very generic layout of a possible tracking system for an LHC experiment.}
\label{fig:dissertori_TrackerLayout}
\end{figure}   
    
    Concerning the intrinsic resolution requirements of the tracking detectors, we can look again
    at equation (\ref{eq:dissertori_Sagitta}) and plug in some typical numbers, such as
    $L=1$ m and $B=4$ T. If we aim for a relative momentum resolution of 1\% at $p=100$ GeV,
    we have to measure the sagitta with a precision of $\delta s = 15 \mu$m. Therefore the 
    individual hit reconstruction precision should be in this ballpark and definitely not much worse 
    than $\mathcal{O}(50-100) \mu$m. Pattern recognition, impact parameter resolution and the
    large soft-particle flux at small $r_\perp$ (cf.\ section \ref{sec:dissertori_LHCparam}) ask for
    the smallest cell sizes and possibly three-dimensional hit reconstruction at the smallest 
    radii. This optimizes the single-hit resolution and minimizes the occupancy and thus fake hit
    assignment. When going to larger radii, the requirements become less stringent, since the
    particle flux falls like $r_\perp^2$ and multiple scattering in the inner layers puts a natural
    limit on the achievable momentum resolution and thus on the necessary cell size (or pitch). 
    On the other hand, the detector area grows
    with $r_\perp^2$, and with it the number of channels and ultimately the costs. A careful optimization
    of all these ingredients has led the two collaborations to go for tracker designs with a basic
    structure as depicted in figure \ref{fig:dissertori_TrackerLayout}. Extremely high-performance
    pixel and silicon strip detectors in the innermost regions, 
    with unprecedented numbers of channels and overall areas, are complemented with a straw-tube
    tracker (ATLAS) or a larger-pitch silicon strip system (CMS) at larger radii. The straw-tube
    tracker gives a very large number of hits and in addition, via transition radiation layers, also provides
    an electron/pion separation. For the all-silicon choice of CMS it is worth noting
    that the outermost tracking layer has a similarly fine pitch like a few layers at intermediate radii.  This
     provides better accuracy at the end of the lever arm than in ATLAS, 
     hence a more than a factor of 2 better momentum resolution at $\eta=0$. 
     The CMS pixel detector is engineered in a manner which allows quick
    and easy installation and removal, a non-negligible feature in view of many years of LHC running
    and the necessary shut-down periods in between. A comparison of the final tracking systems of
    ATLAS and CMS can be found in tables 4 and 6, 
    and of their performance in table 7 of Ref.\ \cite{Froidevaux:2006rg}. In terms of momentum
    resolution the CMS tracker turns out to be superior to that of ATLAS ($1.5\%$ at $p=100$ GeV and
    central rapidity, compared to $3.8\%$), in particular at larger rapidities. This is because of the stronger
    magnetic field and its better uniformity over the whole tracker region. On the other hand, the
    vertexing and b-tagging performances are similar, and the impact of the material and the larger
    $B$-field seem to be visible in slightly lower reconstruction efficiencies (80-85\%) for low-momentum 
    pions and electrons in CMS, compared to ATLAS (84-90\%).  
   
The tracking systems described above play an important role also for the momentum
measurements of muons. However, when designing the muon systems, further
requirements have to be taken into consideration:
\begin{itemize}
\item muon momenta up to 1 TeV should be reconstructed at a precision of 10\%, over a wide
     rapidity range;
\item the mass of high-mass di-muon resonances, such as a hypothetical $Z^\prime$ 
        with mass of $\mathcal{O}(1\, \mathrm{TeV})$, should be reconstructed at a 
        precision of 1\%;
\item muon identification has to be performed in a very dense environment, and if needed
      the systems should be capable of triggering on and measuring muons above 
      $p\sim 5$ GeV in stand-alone mode;
\item muon identification and triggering should be based on robustness and redundancy, with radiation-hard detectors
         and various readout speeds. This can be achieved by combining different technologies
         for the muon chambers.
\end{itemize}

The issue of stand-alone muon reconstruction has already been addressed above. Whereas
the interaction point as additional constraint helps to achieve this in CMS, the multiple
scattering poses some limitations, which are avoided by air-core toroids in ATLAS. The effect
of multiple scattering on the momentum resolution can be modeled by
\begin{equation}
 \label{eq:dissertori_multipleScattering}
   \frac{\delta p_\mathrm{ms}}{p} \approx \frac{52\;10^{-3}}{\beta\,B\,\sqrt{L\,x_0}} \; . 
\end{equation}
If we choose $\beta \sim 1$ for the particle's velocity, $B=2$ T, $L=2$ m and a radiation
length of $x_0=0.14$ m,  as is the case in the 
iron return yoke of CMS, we find a relative uncertainty of $5\%$, which places an absolute
limit on the achievable resolution. Further issues are the unprecedented
challenges to be faced by the alignment systems and the punch-through of pions from the
calorimeters. 

\begin{figure}[htbp]
\centering
\begin{tabular}{c}
\includegraphics[width=9cm]{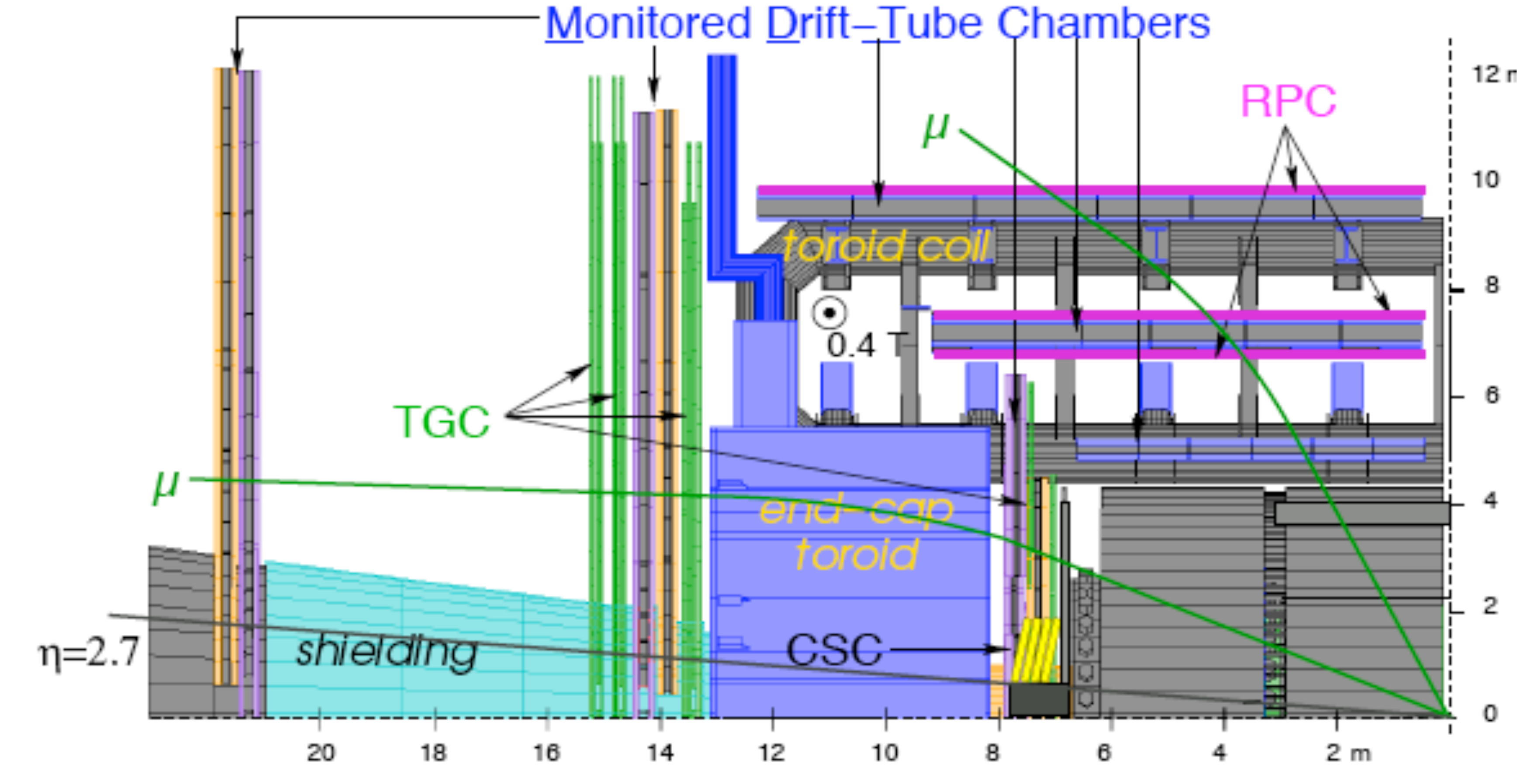} \\[0.3cm]
\includegraphics[width=9cm]{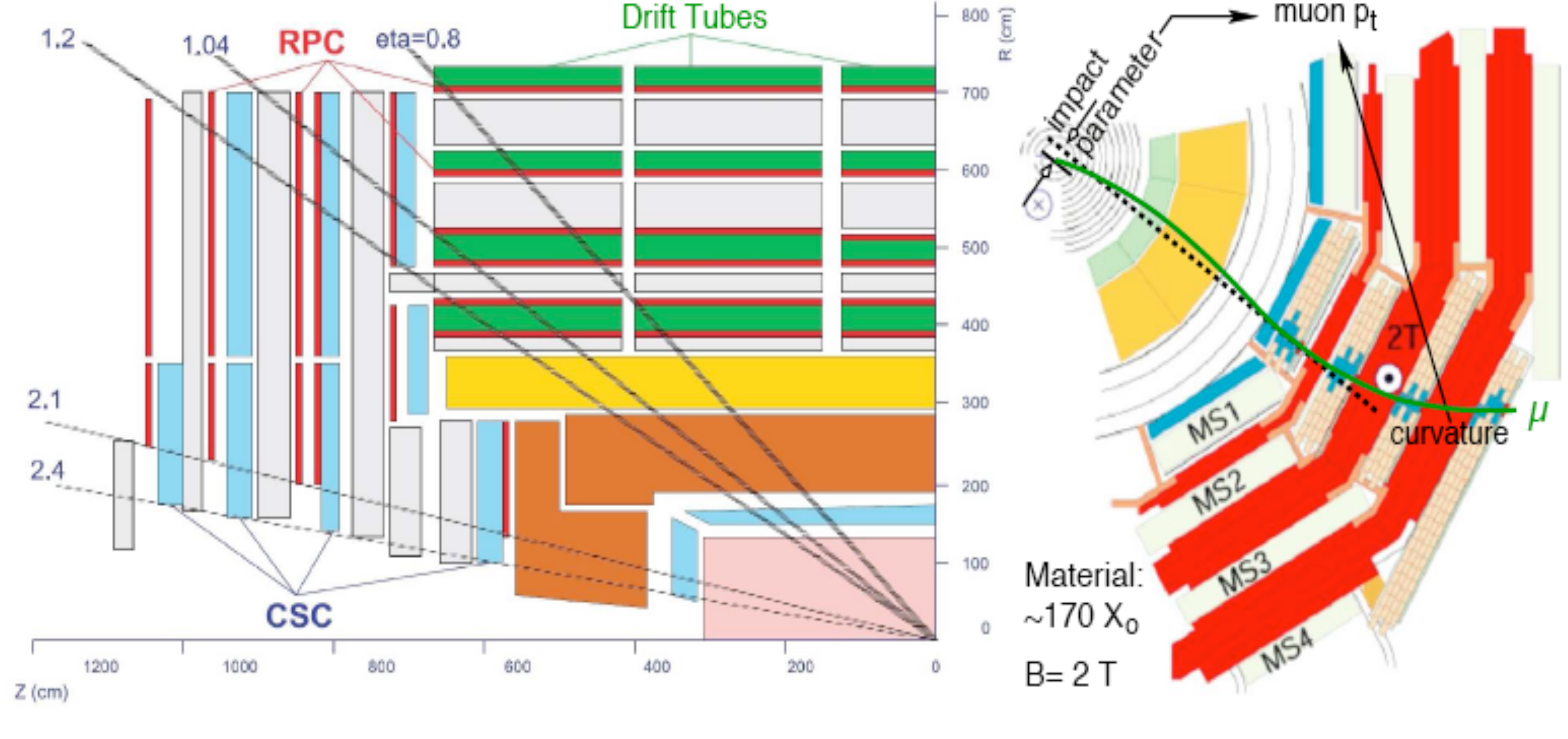} 
\end{tabular}
\caption{Basic layouts of the muons systems in ATLAS (top) and CMS (bottom).}
\label{fig:dissertori_MuonSystems}
\end{figure}

The ATLAS muon spectrometer excels by its stand-alone reconstruction and triggering capabilities 
and its large coverage ($|\eta|<2.7$) in open geometry. At the same time, the complicated geometry and field
configurations lead to large fluctuations in acceptance and performance over the full potential
$\eta-\phi$ area. The CMS muon system allows for a superior momentum resolution in the
central detector regions, when combining the information from the inner tracker with the muon chambers.
This overall excellent resolution is degraded in the forward regions ($|\eta|>2$), where the 
solenoid bending power becomes insufficient. In addition, the limitations on stand-alone triggering
under high-rate conditions have already been discussed. Again, a detailed comparison of the systems
can be found in \cite{Froidevaux:2006rg} (tables 11 and 12), and their basic layouts are depicted
in figure \ref{fig:dissertori_MuonSystems}.

%%%%%%%%%%%%%%%%%%%%%%%%%%%%%%%%%%%%%%
% Calorimeters
%%%%%%%%%%%%%%%%%%%%%%%%%%%%%%%%%%%%%%
\subsection{Calorimetry}
 \label{sec:dissertori_Calo} 

\begin{figure}[htbp]
\centering
\includegraphics[width=11.5cm]{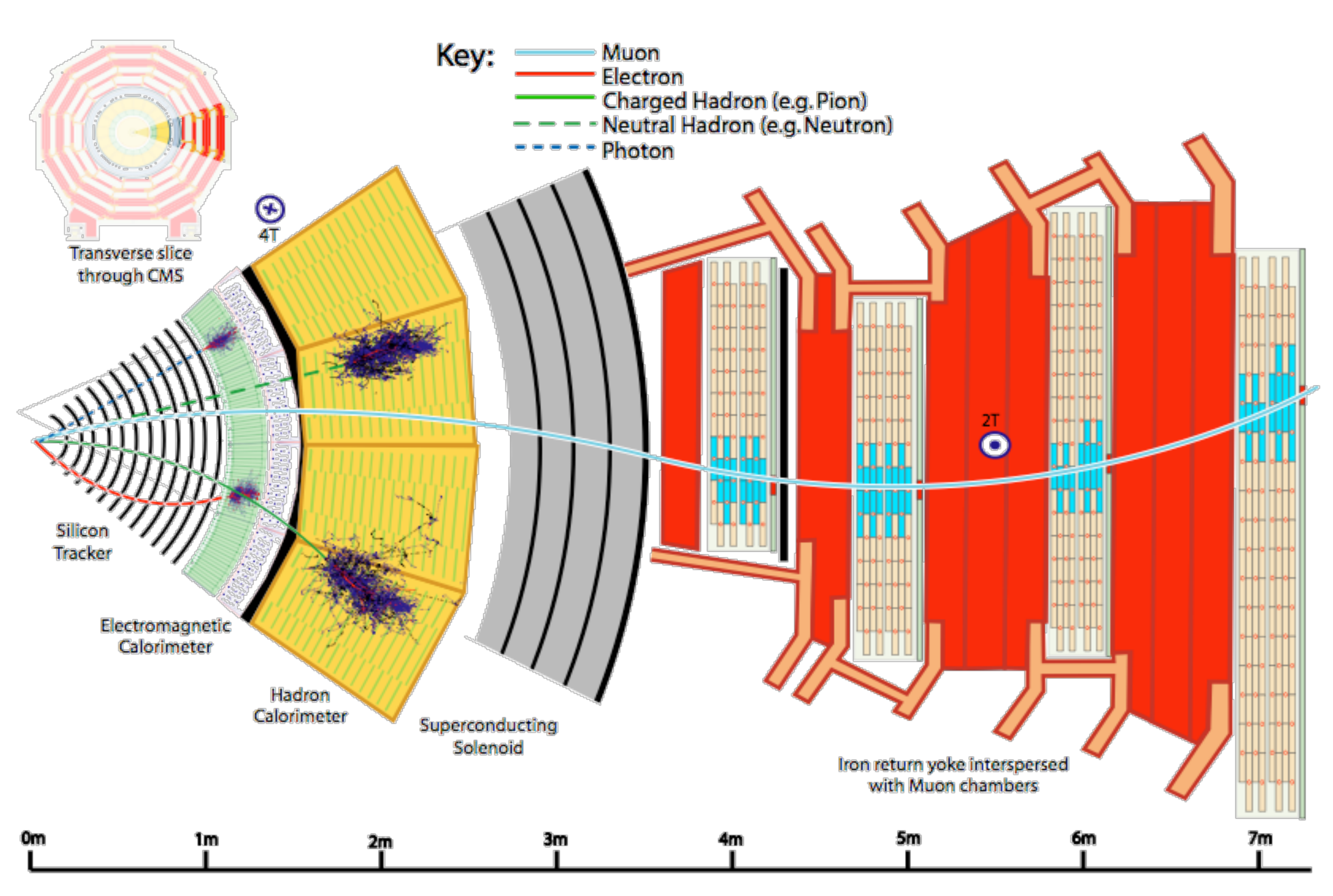} 
\caption{Transverse slice through CMS, from \cite{CMS-slice}.}
\label{fig:dissertori_CMSslice}
\end{figure}

Calorimeters play a central role in the reconstruction and (transverse) energy measurement
of electrons, photons and hadronic jets. Whereas the best energy/momentum resolution for
low- and medium-$\pt$ particles is obtained with spectrometers, at very
high $\pt$ the calorimeters take over, as can be seen from
\begin{equation}
  \label{eq:dissertori_resolution}
   \mbox{Calorimeter:}\;\;\frac{\delta E}{E} \propto \frac{1}{\sqrt{E}} \quad\quad
   \mbox{Spectrometer:}\;\;\frac{\delta p}{p} \propto p \; .
\end{equation}
This is also one of the reasons why calorimeters provide essential
information for triggering on high-$\pt$ objects. At LHC they have to absorb
particles and jets with energies up to the TeV region. This has the following
implications: The position of the maximum of a shower, which develops in a calorimeter,
grows like $\ln E$, namely $x_\mathrm{max} \propto x_0\,\ln(E/E_c)$, where $x_0$ and $E_c$ are the
radiation length and critical energy of an absorber material (see, e.g.\ \cite{Amsler:2008zzb}). 
In order to well contain an electromagnetic
shower of order 1 TeV we need an absorber thickness of $\sim 25 x_0$,  similarly,
for containing a hadronic jet of 1 TeV we need roughly $11 \lambda_0$. Here 
$\lambda_0$ is the effective interaction length of the system. Now let's take
some concrete examples, such as Lead-Tungstate (PbWO$_4$) for the electromagnetic calorimeter (ECAL).
We have $(x_0)_{\mathrm{PbWO}_4} = 0.89$ cm, which implies that for this
ECAL, leaving also some space for the readout electronics, we have to foresee at least
some 50-60 cm of radial space. On the other hand, if we choose iron as absorber material
for the hadronic calorimeter (HCAL), we find $(\lambda_0)_\mathrm{Fe} = 16.8$ cm and thus
require about 180 cm of radial thickness in order to fully contain 1 TeV jets. However,
if we remember the constraints in the CMS design, mostly given by the coil diameter and 
the tracker size, we observe that 
$R_\mathrm{coil} - R_\mathrm{tracker} - \mathrm{ECAL(+electronics)} \sim 1$ m
(cf.\ figure \ref{fig:dissertori_CMSslice}). 
Thus, instead of 11 interaction lengths there is only space for 6, or $7 \lambda_0$ when counting also the ECAL
material in front of HCAL. Indeed, a detailed analysis shows that the CMS coverage,
for $|\eta| \lesssim 1$, is smaller than the qualitative requirement of $11 \lambda_0$. In order
to remedy this situation, a so-called \textit{tail-catcher} (or HO=HCAL-Outer) is installed 
externally to the coil, just before the first muon stations. In the case of ATLAS
no such problem exists. Because of the different magnet system, as discussed above, there
is enough space left between the central solenoid and the external toroids to place
electromagnetic and hadronic calorimeters of sufficient absorption lengths. 

Further issues to be considered when choosing a calorimeter system are~:
\begin{itemize}
 \item homogeneous vs.\ sampling calorimeter: whereas the former typically has the advantage
  of providing a better resolution (especially the stochastic term), a sampling calorimeter
  offers the possibility to measure also the longitudinal shower development at different
  sampling depths;
  \item the very-forward calorimeters (at pseudo-rapidities up to 5) can either be put
   at some larger distance in order to reduce the radiation load, or be kept as close as possible
   to the other calorimeter parts, thus giving better uniformity of the rapidity coverage. A careful
   choice of radiation-hard materials has to be made in this case;
   \item when choosing the projective calorimeter tower sizes the relevant parameters are
    the Moli\`ere radius and the expected and acceptable occupancy. For example, a very simple
    solution for an HCAL segmentation could be 
    $\Delta\eta \times (\Delta\Phi/2\pi) = 0.1 \times 0.1$, which for a complete rapidity coverage
     of $2\,\eta_\mathrm{max} = 10$ would lead to about 10 000 towers with the corresponding
     number of readout channels.
\end{itemize}

ATLAS and CMS have made some quite distinct choices in calorimeter technologies
and layouts. Detailed comparisons and the main design parameters can be found in
tables 8, 9 and 10 of Ref.\ \cite{Froidevaux:2006rg}. A rough comparison is also
given in table \ref{tab:dissertori_ATLAS-CMS-Comparison} below.

%%%%%%%%%%%%%%%%%%%%%%%%%%%%%%%%%%%%%%
% DAQ
%%%%%%%%%%%%%%%%%%%%%%%%%%%%%%%%%%%%%%
\subsection{Data Acquisition (DAQ)}
 \label{sec:dissertori_DAQ} 

When discussing the main LHC parameters in section \ref{sec:dissertori_LHCparam},
we have already encountered some of the relevant numbers, which determine the
design of the multi-level Trigger/DAQ architectures.  The online requirements can be
roughly summarized by a collision rate of 40 MHz, an event size of $\sim 1$ Mbyte, 
a Level-1 Trigger input of 40 MHz, a Level-2 (or High-Level) Trigger input of 100 kHz, a mass storage
rate of $\sim 100$ Hz, thus an overall event rejection power of $\sim 10^7$ with
a system dead-time not exceeding the per-cent level. Further DAQ design issues are
a data network bandwidth (for the event builder) of $\sim$ TByte/sec, a computing power
needed for the High-Level Trigger (HLT) of $\sim 10$ Tflop, corresponding to about 10 000
computing cores and a local storage need of $\sim 300$ TByte. The systems have 
to be robust, i.e.\ the operational efficiency should be independent of detector noise and
machine conditions. In order to estimate the trigger efficiencies in a purely data-driven way
(without Monte Carlo simulations), multiple overlapping triggers have to be carefully designed. 
Finally, also triggers and data streams for alignment and calibration have to be provided.
A guiding principle in order to meet this formidable task was to minimize custom design 
as much as possible, and rather exploit the fast developments in data communication
and computing technologies. Indeed, at the early times some technologies, such as 
network switches, were not yet in the performance range required. However, the decision
to count on and extrapolate the technology advances has paid off in the end.  

\begin{figure}[htbp]
\centering
\includegraphics[width=11.5cm]{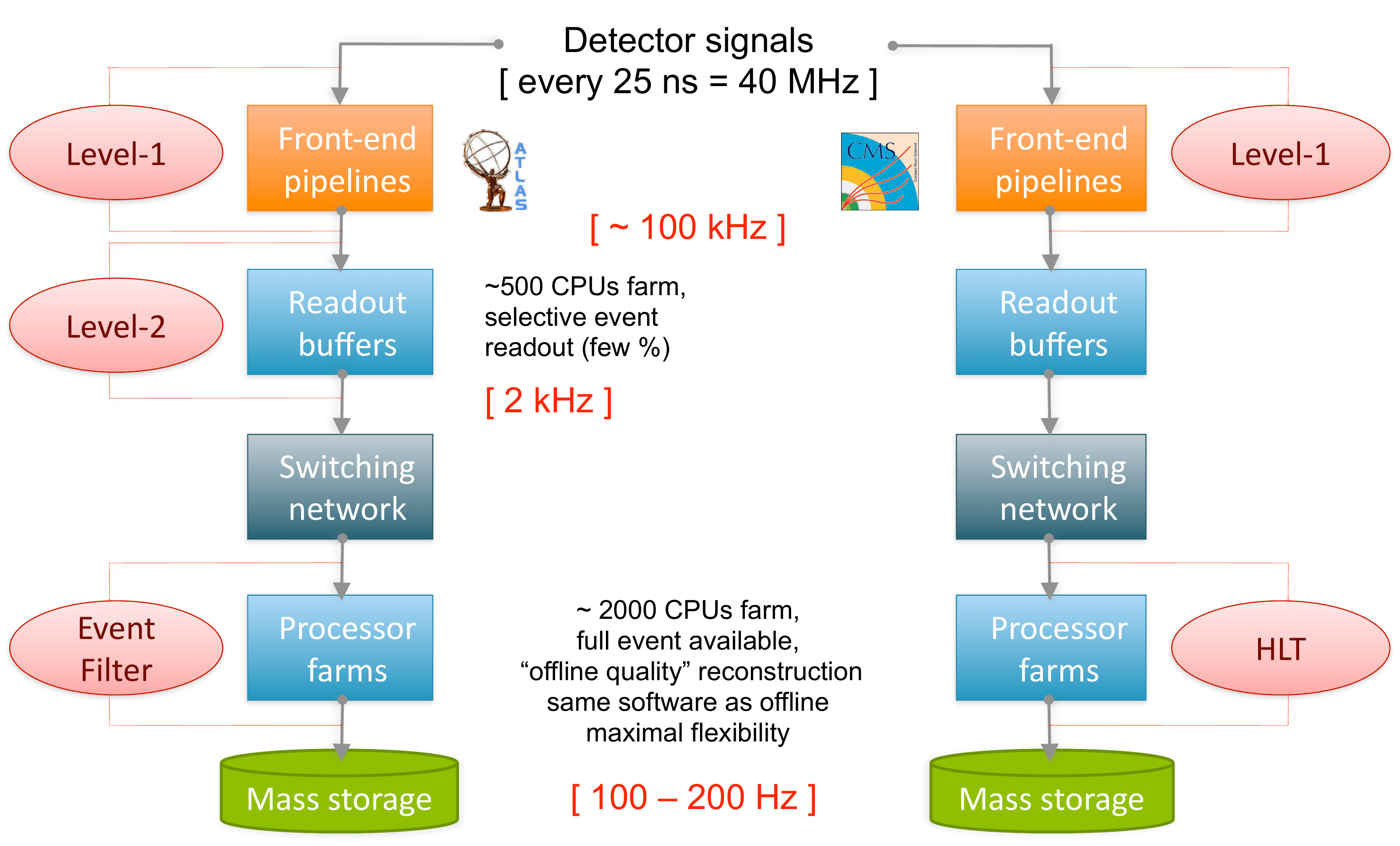} 
\caption{Overview of the trigger and readout schemes in ATLAS and CMS, from \cite{Hoecker-CERN}.}
\label{fig:dissertori_DAQoverview}
\end{figure}

The basic trigger schemes are depicted in figure \ref{fig:dissertori_DAQoverview}. Both
experiments have chosen similar approaches, with some differences however, in particular
regarding the steps from the first level to the high-level trigger. Obviously, in both cases
one starts with a channel data sampling at 40 MHz. The Level-1 trigger has to select
events at a rate of 100 kHz, with decisions based on the identification of (relatively-)high 
$\pt$ electrons, muons, jets, as well as $\;\MET$. This is achieved by local pattern 
recognition and by energy estimates from prompt macro-granular information. The time
budget for taking a decision is $3.2\,\mu$sec, i.e., 128 bunch crossings. This budget includes
the time needed to transfer the signals to the central logic and back, as well as the
time needed by the logic (implemented on custom-designed boards) itself. During this time,
until a Level-1 "Accept" or "Reject" arrives, 
the signals are stored in a pipeline (readout buffers on the front-end boards) of at least
$3\,\mu$sec length. All this requires a high-precision ($\sim 100$ ps) timing, trigger and control
distribution system.

The next level ("Level-2") differs somewhat between the experiments. Whereas in ATLAS it is a
dedicated trigger stage, implemented on a farm of CPU's, in CMS it can be regarded as a first
stage of a more general HLT processing.
 Indeed, in CMS the whole Level-1 output bandwidth has
to be passed through a switching network, into a large farm of CPU's, where the events are
processed in a parallel manner, i.e., one event per processor. At this stage the full event
information is available and a reconstruction of almost ``offline quality'' is possible. This has
the great advantage of full flexibility, in particular if need arises to react to special machine/detector
conditions or to new physics scenarios. The rate reduction from 100 kHz to $\sim 100$ Hz
can be achieved with about 10 000 cores and a time budget of 100 msec/event. The price to
pay for this flexible and scale-free model is the need for an extremely high-performance 
network switch for the data distribution. In ATLAS the switch requirements are less demanding
because of the intermediate Level-2 stage. However, this implies that only a 
very selective event readout (a few \% of the whole event information) is processed here, on the basis of 
so-called regions of interest as identified at Level-1.

After an HLT ``Accept'', the events are fed into data streams which then are 
distributed onto several primary datasets. The latter are typically identified by a set of
trigger bits. These datasets are transferred to the Tier-1 centres for a first reprocessing
step and then further separated into smaller (secondary) datasets or special
data skims, for user analysis at the Tier-2 and Tier-3 computing centres. All
this is based on the LHC Computing Grid concept, explained in a dedicated set of
lectures \cite{Charpentier-lecture}.

%%%%%%%%%%%%%%%%%%%%%%%%%%%%%%%%%%%%%%
% Overview
%%%%%%%%%%%%%%%%%%%%%%%%%%%%%%%%%%%%%%

\section{Overview of ATLAS and CMS}
 \label{sec:dissertori_ATLASCMS} 

\begin{figure}[hbtp]
\centering
\begin{tabular}{c}
\includegraphics[width=11cm]{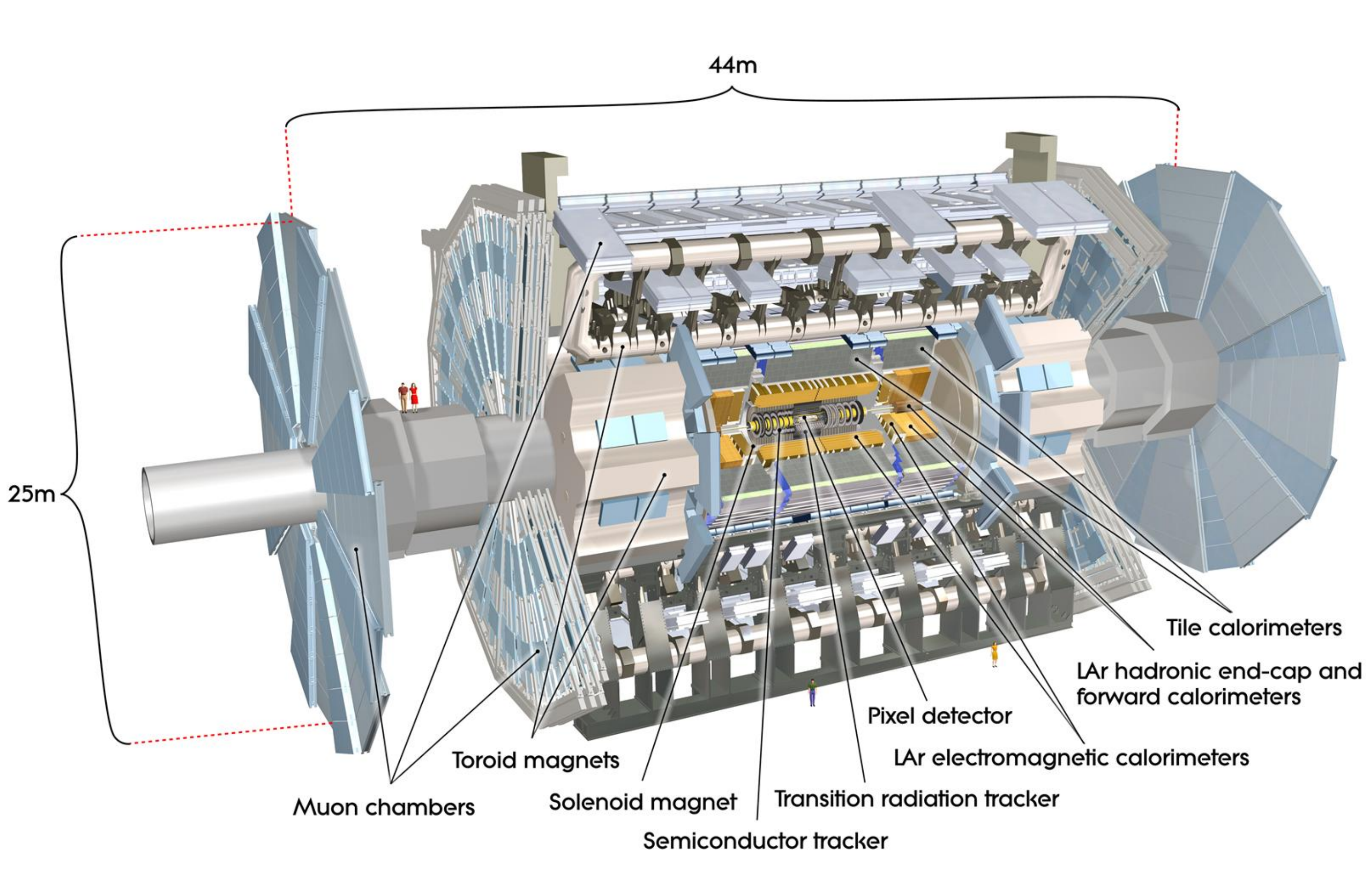} \\[1cm]
\includegraphics[width=11cm]{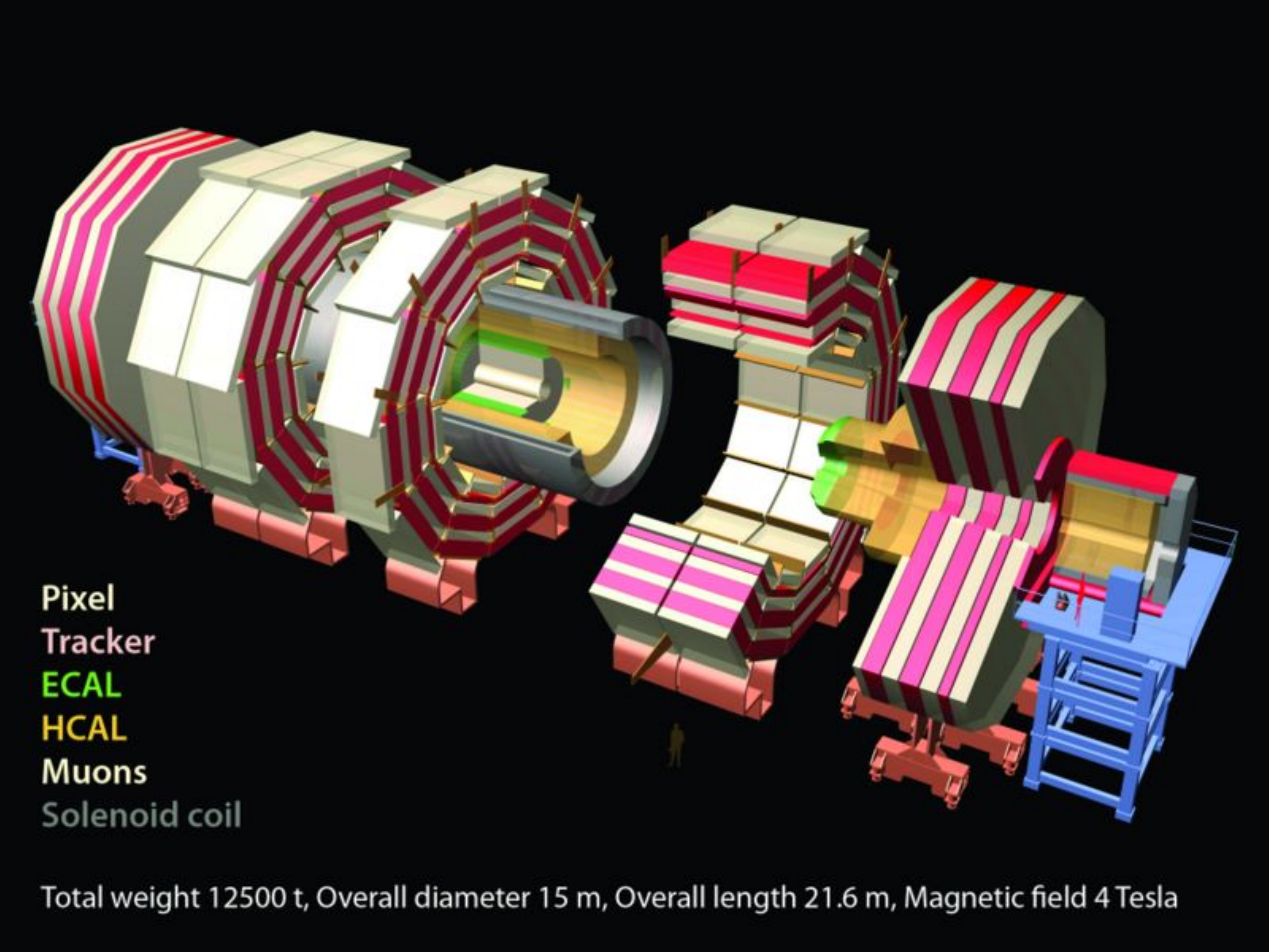} 
\end{tabular}
\caption{General layout and overview of the main elements of ATLAS (top) and CMS (bottom).}
\label{fig:dissertori_ATLAS-CMS}
\end{figure}

All the issues, requirements and boundary conditions, 
as sketched in the sections above, have been taken into account for the
final designs of the two large general-purpose LHC experiments, ATLAS and CMS. 
A number of their most characteristic features have already been hinted at before,
and therefore will not be repeated here. Their general layouts are drawn in
figure \ref{fig:dissertori_ATLAS-CMS}, and a rough comparison of their most
important components and their performance is given in table \ref{tab:dissertori_ATLAS-CMS-Comparison}.
However, this table is only intended to give a quick and rough overview. For a much
more detailed comparison, in particular regarding the expected and so-far observed
performance, the relevant literature should be consulted, such as
\cite{Froidevaux:2006rg} or the set of publications and presentations, 
which have resulted from the commissioning
with cosmic rays and first beams. Here it is also worth noting that the performance in terms
of jet energy and $\MET$ resolution, as expected from  pure calorimeter resolutions,  can be
considerably improved, especially at low $\pt$. This is possible via so-called \emph{Particle Flow} approaches,
where the information from the calorimeters is combined in an optimal way with the input
from the other sub-detector systems, in particular the tracking. Such an algorithm attempts to
reconstruct all particles in an event and thus extract the maximal information from the available data.

\begin{table}[htdp]
\begin{center}
\begin{footnotesize}
\begin{tabular}{|l|l|l|}
\hline
 & & \\[-0.2cm]
 & \textbf{\large{ATLAS}} & \textbf{\large{CMS}} \\
 & \footnotesize{\textbf{A}  \textbf{T}oroidal  \textbf{L}HC  \textbf{A}pparatu\textbf{S}} & 
 \footnotesize{\textbf{C}ompact  \textbf{M}uon  \textbf{S}olenoid} \\
\hline
 & Air-core toroids + & Solenoid \\
 Magnet(s) & solenoid in inner cavity & \\
 & four magnets & only one magnet \\
 & calorimeters in field-free region & calorimeters inside field \\
 \hline
  & Pixels and Si-strips & Pixels and Si-strips \\
  Tracker & Pid: TRT +  $dE/dx$ & Pid: $dE/dx$ \\
  & $B=2$ T & $B=3.8$ T \\
  & $\sigma_{\pt}/\pt \sim 5\times 10^{-4} \pt \oplus 0.01$ & 
     $\sigma_{\pt}/\pt \sim 1.5\times 10^{-4} \pt \oplus 0.005$ \\
  \hline 
 & Lead-liquid argon & Lead-Tungstate crystals \\  
 Electromagnetic & sampling & homogenous \\ 
 Calorimetry & $\sigma_{E}/E \sim 10\%/\sqrt{E} \oplus 0.007$ &
 			 $\sigma_{E}/E \sim 3\%/\sqrt{E} \oplus 0.5\%$  \\
  & longitudinal segmentation & no longitudinal segmentation \\
  \hline
  & Fe-scint.\ + Cu-liquid argon & Brass-scint. \\
Hadronic  & $\sigma_{E}/E \sim 50\%/\sqrt{E} \oplus 0.03$ &
 			 $\sigma_{E}/E \sim 100\%/\sqrt{E} \oplus 0.05$  \\
 Calorimetry &  $\gsim 11 \lambda_0$ &  $\gsim 7 \lambda_0$ + tail catcher \\
   & longit.\ segmented readout & single (full) depth in readout \\
 \hline
              & combined with air-core toroids & instrumented iron return yoke \\	
 Muon & $\sigma_{\pt}/\pt \sim 2\%\; (\mathrm{at}\, 50\, \mathrm{GeV}) $& 
               $\sigma_{\pt}/\pt \sim 1\%\; (\mathrm{at}\, 50\, \mathrm{GeV}) $ \\
  System &  $\quad\quad\quad\;\;\,\sim 10\%\; (\mathrm{at}\, 1\, \mathrm{TeV})$  & 
                    $\quad\quad\quad\;\;\,\sim 10\%\; (\mathrm{at}\, 1\, \mathrm{TeV})$ \\
                    & in stand-alone mode & when combined with tracker \\
                    \hline
\end{tabular}
\end{footnotesize}
\caption{A simple comparison of the main design choices and performance numbers for
 the ATLAS and CMS detectors. Pid=Particle identification. The calorimeter energy resolutions are
 for the barrel parts. For considerably more detailed information and discussions
 the Refs.\ \cite{:2008zzm,:2008zzk,Froidevaux:2006rg} 
 should be consulted. \label{tab:dissertori_ATLAS-CMS-Comparison}}
\end{center}
\label{default}
\end{table}%

There is one more design feature, which has not been discussed before and which
distinguishes the two experiments quite considerably. CMS is strongly characterized by
a very modular design, which has been a guiding principle from its conception onwards. 
It was clear that it was not possible to build 13 m long muon chambers, which would
cover the whole barrel. This led to the idea of separating the barrel into 5 completely
independent wheels, with the central one supporting the magnet coil. This obviously
offers great flexibility in construction and maintenance, in particular since it was anticipated
to construct and test the experiment at the surface, in parallel with the cavern excavation.
Indeed, CMS was lowered into the cavern by a sequence of heavy-lifting operations of
it's individual elements. In the cavern the modular structure has the further advantage
of easy access to sub-parts of the detectors during shutdown periods. The most dramatic
example is the pixel detector, which can be removed and re-installed relatively quickly.
ATLAS has been assembled inside its cavern, and access to some of the 
inner elements of the detector poses a greater challenge and requires more time than in CMS.

%%%%%%%%%%%%%%%%%%%%%%%%%%%%%%%%%%%%%%
% commissioning
%%%%%%%%%%%%%%%%%%%%%%%%%%%%%%%%%%%%%%
\subsection{From construction to first collisions}
 \label{sec:dissertori_commissioning} 

The numerous quality checks during 
construction and beam tests of series detector modules let us conclude that the detectors as built should
give a good starting-point performance. However, commissioning of the detectors with
cosmic rays, beam-splash events and first collisions has been and is still an invaluable tool
to prepare the experiments for the high-energy operations. The beam-splash events, which
were available already in 2008 during the very first LHC injection tests, turned out to
be extremely useful to time-in many subsystems and to align and intercalibrate some of the
subdetectors. Such splash events were observed when the 450 GeV injection beam, with
about $10^9$ protons in a single bunch, struck collimators some 150 m upstream of the 
experiments, giving $\sim 10^5$ muons traversing the detectors, mostly horizontal. Halo muons
were also observed once the beam started passing through the detectors. These very high-energy
particles give almost straight tracks across the systems, and thus can be used for alignment studies.

\begin{figure}[htbp]
\centering
\begin{tabular}{lr}
\includegraphics[width=6cm]{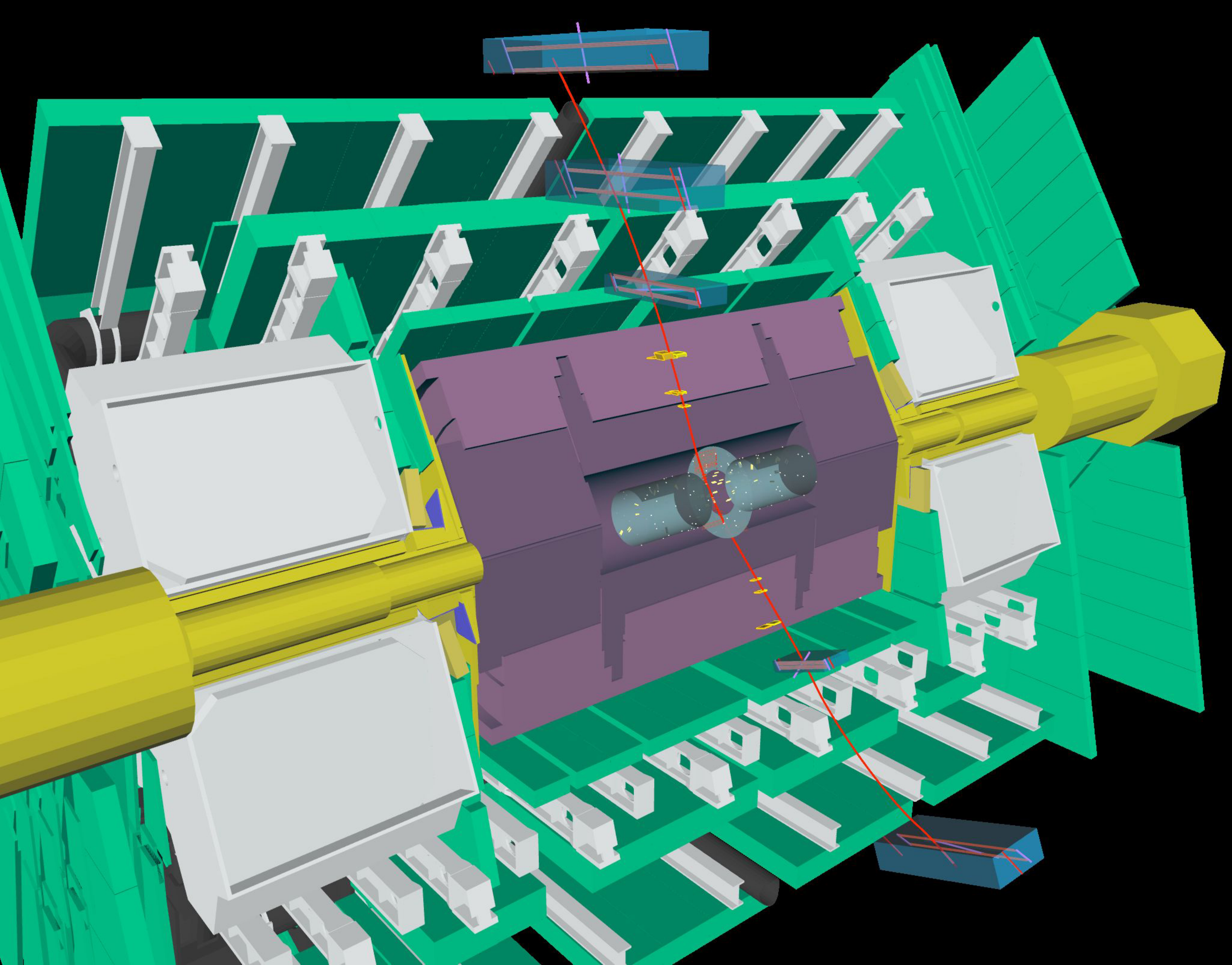} &
\includegraphics[width=5cm]{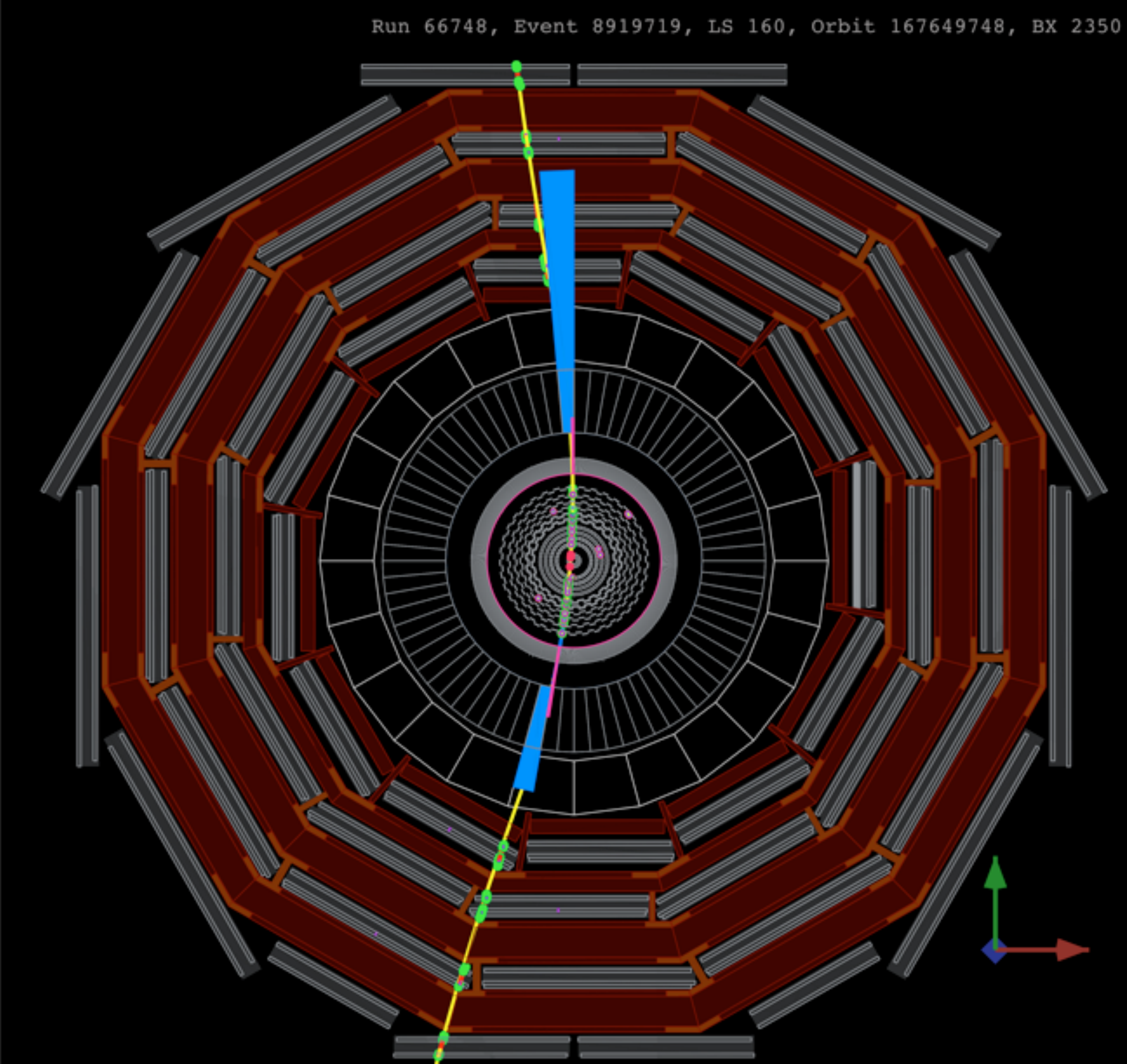} 
\end{tabular}
\caption{Event displays of cosmic rays in ATLAS (left) and CMS (right).}
\label{fig:dissertori_EvdispCosmics}
\end{figure}

Before the first LHC start, and between the incident on Sep 19, 2008 and the re-start in Nov 2009,
all the LHC experiments made extensive use of cosmic rays. Besides setting up the online operations,
bringing all subsystems into a unified readout and training the data-taking procedures as far as possible,
the cosmic runs gave an astonishing number of commissioning, calibration and even some
physics results. Each experiment collected several hundred million of cosmic events (cf.
 figure \ref{fig:dissertori_EvdispCosmics}), leading up to
$\sim 1$ PByte of raw data. Track reconstruction in the muon and tracker systems, in stand-alone
mode or combined, could be exercised and its performance, such as the momentum resolution, 
measured. The modeling of the magnetic field maps could be verified and corrected where necessary.
Energy deposits by muons in the calorimeters were registered and compared to 
predictions. All this has helped the experiments to approach the first collision period in an
extremely well prepared fashion, with often more than 99\% of the subdetector channels fully
functional and well understood. Obviously, with the first collisions in hand, 
the trigger and data acquisition systems were finally timed-in,
the data coherence checked, sub-systems synchronized and reconstruction algorithms debugged
and calibrated.

Indeed, the speed, at which the experiments analyzed the first collision data
% (cf.\ figure \ref{fig:dissertori_Evdisp_firstcoll}) 
at the end of 2009, as well as the quality of the results and the
agreement of the data with the Monte Carlo predictions was a surprise to many.
This also gives strong hopes that the high-energy collisions starting in 2010 will
lead to high-quality data, analyzed in a timely fashion.  

%\begin{figure}[htbp]
%\centering
%\begin{tabular}{c}
%\includegraphics[width=9cm]{ATLAS_Evdisp_FirstColl} \\
%\includegraphics[width=9cm]{CMS_Evdisp_Multijet} 
%\end{tabular}
%\caption{Event displays of the first collisions registered in ATLAS (top) and CMS (bottom) on
% Nov 23, 2009, at a beam energy of 450 GeV.}
%\label{fig:dissertori_Evdisp_firstcoll}
%\end{figure}

Nevertheless, considerable effort will have to be invested for obtaining the ultimate
calibration and alignment precision.
The electromagnetic and hadronic calorimeters will be calibrated with
physics events. For example, the initial crystal inter-calibration precision of about 4\%
for the CMS ECAL will be improved to about 2\% by using the $\phi$-symmetry of the energy
deposition in minimum-bias and jet events.
Later the ultimate precision ($\approx 0.5\%$) and the absolute
calibration will be obtained using $Z\rightarrow e^+ e^-$ decays and
the $E/p$ measurements for isolated electrons, such as in $W\rightarrow e\nu$ decays
\cite{Bayatian:2006zz}.
The latter requires a well understood tracking system. The uniformity of the 
hadronic calorimeters can be checked with single pions and jets. In order to obtain the
jet energy scale (JES) to a few per-cent precision or better, physics processes such as 
$\gamma + \mathrm{jet}$, 
$Z(\rightarrow \ell\ell) + \mathrm{jet}$ or $W\rightarrow$ 2 jets in top pair events
will be analyzed.
Finally, the tracker and muon system alignment will be carried out with generic tracks, isolated muons or
$Z\rightarrow \mu^+ \mu^-$ decays. Regarding all these calibration
and alignment efforts, the ultimate statistical precision should be 
achieved very quickly in many cases. Then systematic effects
have to be faced.

%%%%%%%%%%%%%%%%%%%%%%%%%%%%%%%%%%%%%%
% Hard processes
%%%%%%%%%%%%%%%%%%%%%%%%%%%%%%%%%%%%%%

\section{Measurements of Hard Processes}
 \label{sec:dissertori_HardProc} 

Before entering the discovery regime, considerable efforts will be
invested in the measurements of SM processes. 
They will serve as a proof for a working detector 
(a necessary requirement before any claim of discovery is made).
Indeed, some of the SM processes are also excellent tools to calibrate
parts of the detector. However, such measurements are also interesting in their own right. 
We will be able to challenge the SM predictions at unprecedented energy and momentum transfer scales,
by measuring cross sections and event features for minimum-bias events, jet production, 
$W$ and $Z$ production with their leptonic decays, as well as top quark production.
This will allow to check the validity of the Monte Carlo generators, both at the highest energy scales 
and at small momentum transfers, such as in models for the omnipresent underlying event. 
The parton distribution functions (pdfs) can be further constrained or measured for the first time in
kinematic ranges not accessible at HERA. Important tools for pdf studies will be 
jet+photon production or Drell-Yan processes. Finally, SM processes such as $W/Z$+jets, multi-jet and top pair 
production will be important backgrounds to a large number of searches for new physics and therefore have to
be understood in detail. 

The very early goals to be pursued by the experiments, once the first data are on tape, are
three-fold~: (a) It will be of utmost importance to commission and calibrate the detectors
in situ, with physics processes as outlined below. The trigger performance has to be understood
in a possibly unbiased manner, by analyzing the trigger rates of minimum-bias events,
jet events for various thresholds, single and di-lepton as well as single and di-photon
events. (b) It will be necessary to measure the main SM processes and 
(c)  prepare the road for possible discoveries. It is instructive to recall the 
event statistics collected for different types of processes. 
For an integrated luminosity of $100\,\mathrm{pb}^{-1}$ per experiment, we expect about $10^6$
$W\rightarrow e\nu$ events on tape, a factor of ten less $Z\rightarrow e^+ e^-$ 
and some $10^4$ $t\bar t\rightarrow \mu + X$ events. If a trigger bandwidth of about 10\%
is assumed for  QCD jets with transverse momentum $\pt > 150$ GeV,  
$b\bar b\rightarrow \mu + X$ and minimum-bias events, 
we will write about $10^5$ events to tape, for each of these channels. 
This means that the statistical uncertainties
will be negligible relatively quickly, for most of the physics cases. The analysis results will be dominated
by systematic uncertainties, be it the detailed understanding of the detector response, theoretical
uncertainties or the uncertainty from the luminosity measurements.

The anticipated detector performance leads to the following estimates for the reconstruction
precision of the most important physics objects:
Isolated electrons and photons can be reconstructed with a relative energy resolution characterized
	         by a stochastic term  of a few per-cent and an aimed-for
	         0.5\% constant term. Typically isolation requirements are defined by putting a cone 
	         around the electron/photon and counting the additional electromagnetic and hadronic energy
	         and/or track transverse momentum within this cone. The optimal cone size in $\eta-\phi$ 
	         space depends on the particular analysis and event topology. For typical acceptance cuts, 
	         such as a transverse momentum above 10-20 GeV and $|\eta| < 2.5$, electrons and photons
	         can be expected to be reconstructed with excellent angular resolution, high efficiency ($\ge 80-90\%$)
	           and small backgrounds. Again, the precise values depend very much on the final state topology and the
	         corresponding tightness of the selection cuts. Most importantly, the systematic uncertainty 
	         on the reconstruction efficiency should be controllable at the 1-2\% level, using in-situ measurements
	         such $Z\rightarrow e^+ e^-$ decays, 
	         with one of the electrons serving as tag lepton 
	         and the other one as probe object for which the efficiency is determined.      
	             
	             Isolated muons, with similar acceptance cuts as mentioned above for electrons, should be 
	         reconstructed with a relative transverse momentum resolution of 1 - 5\% and excellent angular
	           resolution up to several hundreds of GeV. Again, a systematic uncertainty 
   	         on the reconstruction efficiency of 1-2\% appears to be achievable.
	         
 Hadronic jets will be reconstructed up to pseudo-rapidities of 4.5 - 5, with good angular resolution.
	         The energy resolution depends rather strongly on the specific calorimeter performance. 
	            For example, in the case of ATLAS (CMS) a stochastic term of the order of 
	            50 - 60\% (100 - 150\%) is to be expected when energy deposits in
	           projective calorimeter towers
	            are used for the jet clustering procedure. However, as mentioned above,
	            important improvements on the CMS jet energy 
	            resolution are expected from new approaches such as particle flow algorithms. 
	            Well above the trigger thresholds jets
	            will be reconstructed with very high efficiency; the challenge is the understanding
	            of the efficiency turn-on curves. In contrast to leptons, for jets the experimental 
	             systematic uncertainties are much more sizeable and difficult to control. A more detailed discussion 
	             will follow below. 	             
	             A further important question is the lowest $\pt$ threshold above which jets
	              can be reconstructed reliably. Contrary to the naive expectation that only high-$\pt$ objects
	              (around 100 GeV and higher) are relevant, it turns out that many physics channels require jets
	               to be reconstructed with rather low transverse momentum of $\sim 20 - 30$ GeV. One reason for this is
	               the importance of jet veto requirements in searches for new physics, such as in the
	               $H\rightarrow WW^*\rightarrow2\ell\,2\nu$ channel, where a jet
	               veto is necessary to reduce the top background. The experimental difficulties related to the 
	               understanding of the low-$\pt$ jet response\footnote{The jet response is defined as the
	                ratio of the reconstructed  and the "true" jet momentum.}, the thresholds due to 
	                noise suppression, the impact of the underlying event and additional pile-up events and
	                ultimately the knowledge of the JES lead to the conclusion that it will
	                 be extremely challenging to reliably reconstruct jets below a $\pt$ of 30 GeV.
	                In addition, also the theoretical predictions are challenged by very low-$\pt$ effects, as 
	                 for example induced by jet veto requirements. Here fixed-order calculations may have to be
	                 supplemented by resummations of large logarithms.

Finally, the missing transverse energy will be a very important ``indirect''
	                 observable, which is constructed from measurements of other quantities, such as
	                  all calorimeter energy deposits. Many searches for new physics, such as Supersymmetry,
	                   rely very much on this observable. However, it turns out that 
	                    it is also an extremely difficult quantity to measure, since it is sensitive to almost
	              every detail of the detector performance. Here it is even more difficult to give estimates
	                of the expected systematic uncertainties. Also, the reconstruction performance depends
	                very much on the details of the particular final state, such as the number of  jets and/or leptons
	                 in the event, the existence of ``true'' missing energy, e.g.\ from neutrinos, the amount
	                  of pile-up events and in general the overall transverse energy deposited in the detector.
	                  The first data are of paramount importance for a timely understanding of this quantity.

In the following I will concentrate on some of the early
 measurements to be performed on the first few hundred pb$^{-1}$ up to 1 fb$^{-1}$ of integrated
luminosity. Many reviews exist on this topic, 
such as Refs.\ \cite{Kane:2008zz,Gianotti:2005fm,Sphicas:2003sb} to mention only a few.
However, before entering the discussion of physics measurements, it is worth recalling some recent
developments in the area of jet algorithms, which will play an important role in almost all of the
LHC analyses.

%Most of the results presented here are taken from the CMS PTDR Vol.\ 2 \cite{Ball:2007zza}, because it
%represents the most recent comprehensive overview compiled by one of the LHC experiments. 

%%%%%%%%%%%%%%%%%%%%%%%%%%%%%%%%%%%%%%
% Jet Algos
%%%%%%%%%%%%%%%%%%%%%%%%%%%%%%%%%%%%%%
\subsection{Jet Algorithms}
 \label{sec:dissertori_JetAlgo} 

In hard interactions, final-state partons and hadrons appear
predominantly in collimated bunches. These bunches are generically
called \textit{jets}.
To a first approximation, a jet can be thought of as a hard parton that
has undergone soft and collinear showering and then hadronization.
Jets are used both for testing our understanding and predictions of
high-energy QCD processes and also for identifying the hard partonic
structure of decays of massive particles like top quarks.
In order to map observed hadrons onto a set of jets one uses a \emph{jet
  definition}. 
Good jet definitions are infrared and collinear safe, simple to use in
theoretical and experimental contexts, applicable to any type of
inputs (parton or hadron momenta, charged particle tracks and/or
energy deposits in the detectors) and lead to jets that are not too
sensitive to non-perturbative effects.
An extensive treatment of the topic of jet definitions is given in
\cite{Moretti:1998qx} (for $e^+e^-$ collisions) and
\cite{Salam:2009jx,Ellis:2007ib} (for $pp$ or $p \bar p$ collisions).
Here I will briefly discuss the two main classes: cone algorithms,
extensively used at hadron colliders, and sequential recombination
algorithms, more widespread in $e^+e^-$ and $ep$ colliders.

Very generically, most (iterative) cone algorithms start with some
seed particle $i$, sum the momenta of all particles $j$ within a cone
of opening-angle $R$, typically defined in terms of (pseudo)-rapidity
and azimuthal angle. They then take the direction of this sum as a new
seed, repeat until the cone is stable and call the contents of the
resulting stable cone a jet if its transverse momentum is above some
threshold $p_{T,\mathrm{min}}$.
The parameters $R$ and $p_{T,\mathrm{min}}$ should be chosen according to the
needs of a given analysis.

There are many variants of cone algorithm, and they differ in the set
of seeds they use and the manner in which they ensure a one-to-one
mapping of particles to jets, given that two stable cones may share
particles (``overlap'').
The use of seed particles is a problem w.r.t.\ infrared and collinear
safety, and seeded algorithms are generally not compatible with
higher-order (or sometimes even leading-order) QCD calculations,
especially in multi-jet contexts, as well as potentially subject to
large non-perturbative corrections and instabilities.
Seeded algorithms (JetCLU, MidPoint, and various other
experiment-specific iterative cone algorithms) are therefore to be
deprecated.
A modern alternative is to use a seedless variant,
SISCone~\cite{Salam:2007xv}.

Sequential recombination algorithms at hadron colliders (and in DIS)
are characterized by a distance $d_{ij} = \min(k^{2p}_{t,i},
k^{2p}_{t,j}) \Delta^2_{ij} / R^2$ between all pairs of particles
$i,j$, where $\Delta_{ij}$ is their distance in the rapidity-azimuthal
plane, $k_{t,i}$ is the transverse momentum w.r.t.\  the incoming
beams and $R$ is a free parameter.
They also involve a ``beam'' distance $d_{iB} = k_{t,i}^{2p}$.
One identifies the smallest of all the $d_{ij}$ and $d_{iB}$ and if it
is a $d_{ij}$ then $i$ and $j$ are merged into a new pseudo-particle
(with some prescription, a recombination scheme, for the definition of
the merged four-momentum).
If the smallest distance is a $d_{iB}$ then $i$ is removed from the
list of particles and called a jet.
As with cone algorithms one usually considers only jets above some
transverse-momentum threshold $p_{T,\mathrm{min}}$.
The parameter $p$ determines the kind of algorithm: $p=1$ corresponds
to the (\textit{inclusive-})$k_t$ algorithm
\cite{Catani:1991hj,Catani:1993hr,Ellis:1993tq},
$p=0$ defines the
\textit{Cambridge-Aachen} algorithm
\cite{Dokshitzer:1997in,Wobisch:1998wt}, while for $p=-1$ we have the
\textit{anti-}$k_t$ algorithm \cite{Cacciari:2008gp}. All these
variants are infrared and collinear safe to all orders of perturbation
theory. Whereas the former two lead to irregularly shaped jet
boundaries, the latter results in cone-like boundaries.
 
Efficient implementations of the above algorithms are available through
the \textit{FastJet} package \cite{Cacciari:2005hq}, which is also
packaged within \emph{SpartyJet} \cite{SpartyJet}.

%%%%%%%%%%%%%%%%%%%%%%%%%%%%%%%%%%%%%%
% QCD Jet production
%%%%%%%%%%%%%%%%%%%%%%%%%%%%%%%%%%%%%%
\subsection{QCD Jet Production}
 \label{sec:dissertori_QCDjets} 

Because of its extremely large cross section, the inclusive
dijet production ($pp \rightarrow 2$ jets + anything) 
completely dominates over all other expected LHC processes with large momentum transfer.
At lowest order in perturbative QCD,
it is described as a $2 \rightarrow 2$ scattering of partons (quarks and gluons),
with only partons in the initial, intermediate and final state. 
Depending on the exchanged transverse momentum (or generally the
energy scale of the scattering process), the final state will consist of
more or less energetic jets which arise from the fragmentation of the outgoing partons.

%Here I will not discuss further this class of measurements, but rather concentrate
%on the parton scattering at large transverse momentum. Examples of envisaged studies of
%minimum-bias events can be found in \cite{Ball:2007zza}.

\begin{figure}[htb]
\begin{center}
\begin{tabular}{cc}
\includegraphics[width=0.45\textwidth]{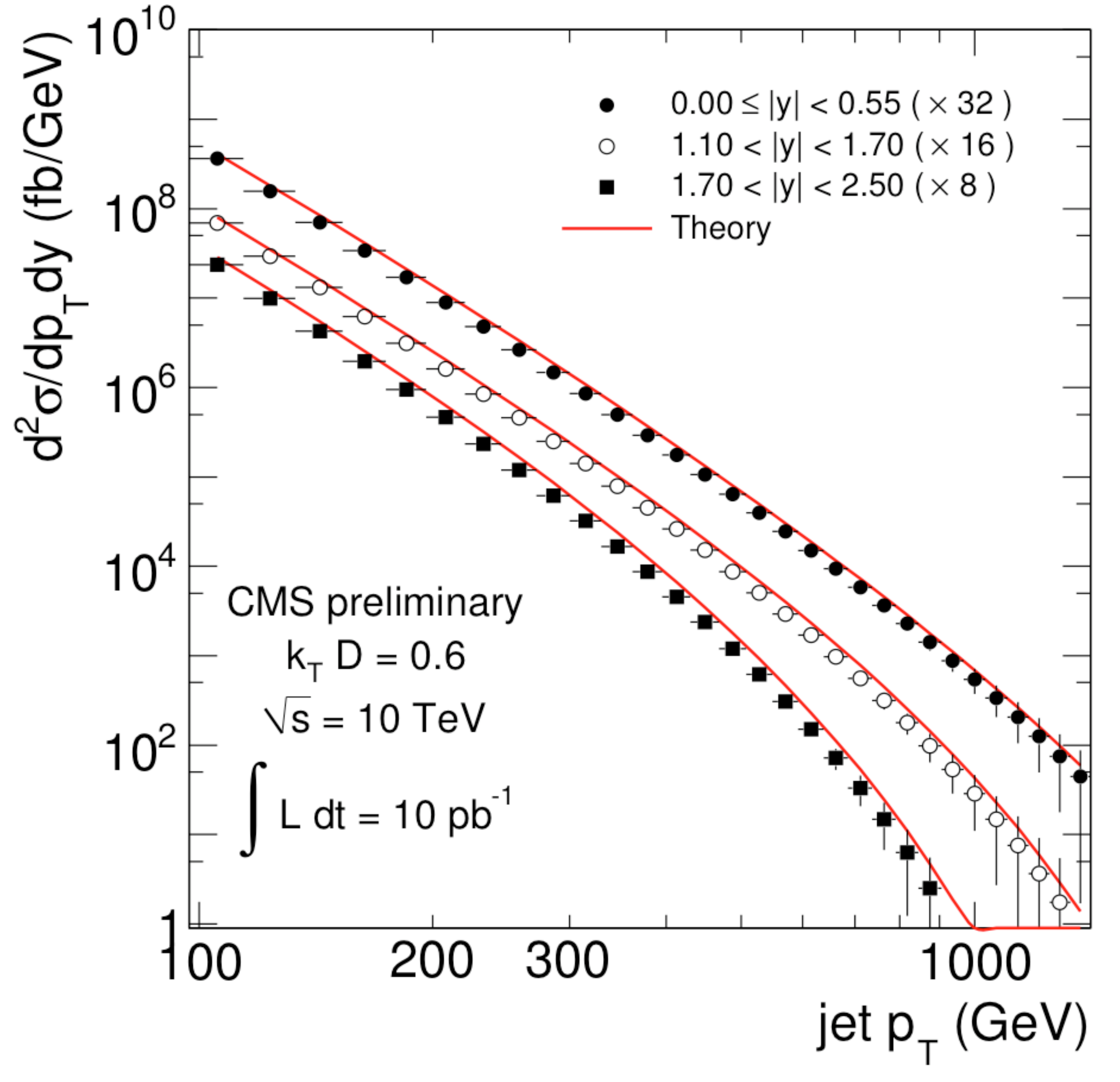} &
\includegraphics[width=0.45\textwidth]{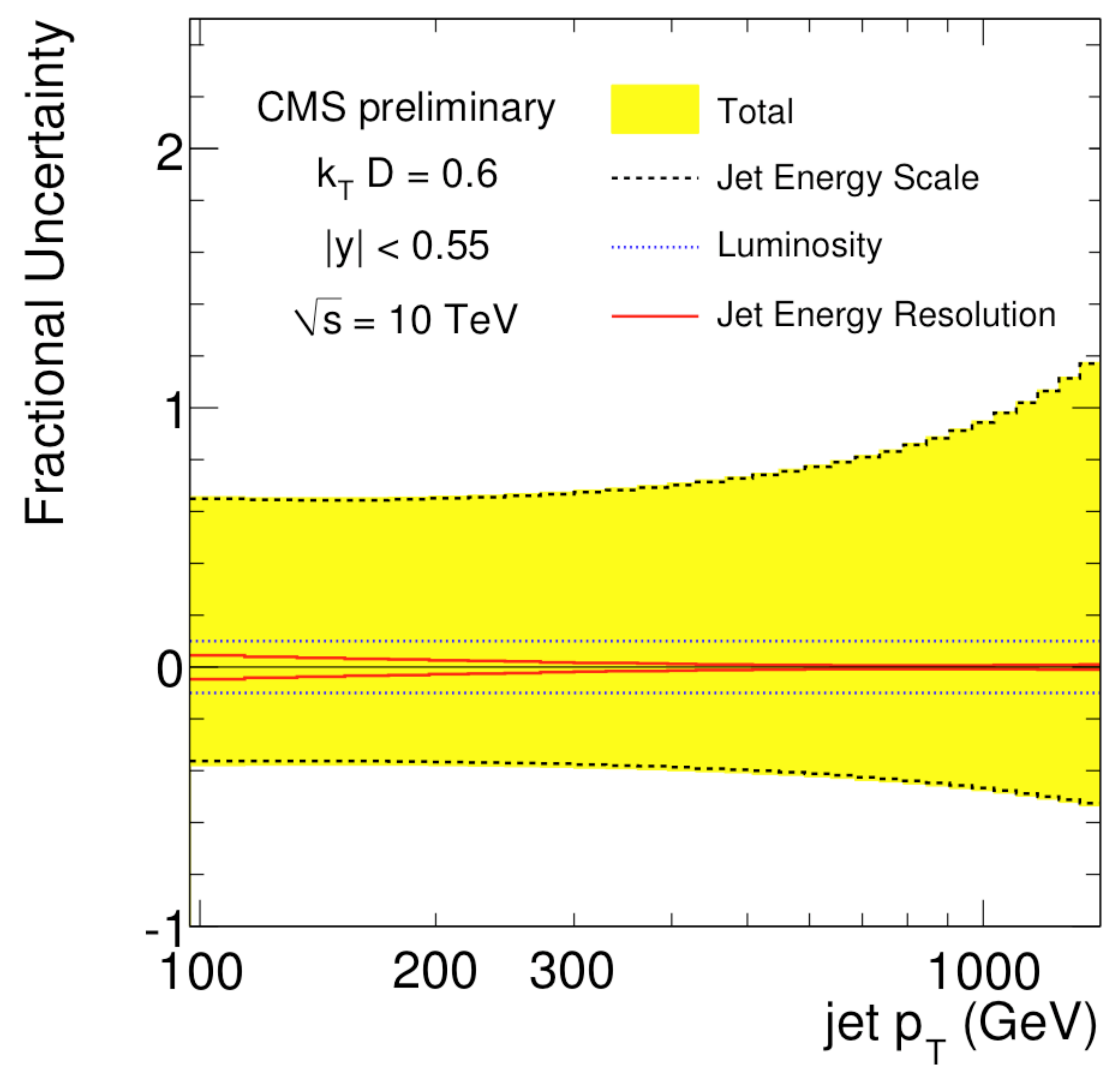} 
\end{tabular}
 \caption{Inclusive jet cross section measurements (left) and related systematic uncertainties (right),
  as foreseen by CMS \cite{CMS-PAS_QCD}.
  }
\label{fig:dissertori_jetxsec}
\end{center}
\end{figure}

For the measurement of the inclusive jet cross section we simply count the 
number of jets inside a fixed pseudo-rapidity  region as a function of jet $\pt$. 
For a second typical measurement, the dijet cross section, events are selected in which the two highest 
$\pt$ jets, the leading jets, are both inside a specified pseudo-rapidity region and 
counted as a function of the dijet (invariant) mass.  Both cases are inclusive processes
dominated by the $2\rightarrow 2$ QCD scattering of partons. The distinction between 
inclusive jets and dijets is only in a different way of measuring the same process. For 
a common choice of the $\eta$ region, events selected by the dijet analysis 
are a subset of the events selected by the inclusive jet analysis, but the number of events in the two analyses coming from QCD is expected to be close at  high $\pt$. The steeply falling cross sections 
are shown in figure \ref{fig:dissertori_jetxsec} (left). For the inclusive jet case, the spectrum roughly follows a 
power law, however, with increasing power for increasing $\pt$, i.e.,
the power increases from about 6 at $\pt=150$ GeV  to about 13 at $\pt = 3$ TeV
and keeps on increasing with jet $\pt$.

Even for very small integrated luminosities the statistical uncertainties will be
negligible, up to very high jet momenta. Thus the Tevatron reach in terms of highest
momenta and therefore sensitivity to new physics, such as contact interactions or heavy resonances,
will be quickly surpassed. For 1 fb$^{-1}$, the inclusive cross section
for central jet production (i.e.\ jet pseudo-rapidities below $\sim 1$) will be known statistically to
better than 1\% up to a $\pt$ of 1 TeV, and the statistical errors on the dijet cross section will
be below 5\% up to dijet masses of 3 TeV.

The real challenge for these measurements will be the determination and control of the jet energy
scale. As mentioned above, the cross sections are steeply falling as a function of jet $\pt$. 
Therefore any relative uncertainty on the jet $\pt$ will translate
into a $n$-times larger relative uncertainty on the cross section, where $n$ indicates the power 
of the spectrum in a specified $\pt$ region, i.e. $d\sigma/d\pt \propto \pt^{-n}$. For example, 
a $5\%$ uncertainty on the energy scale for jets around 100-200 GeV of transverse momentum 
induces a $30\%$ uncertainty on the inclusive jet cross section. This is shown in 
Fig.\ \ref{fig:dissertori_jetxsec} (right), here for the case of a 10\% JES uncertainty. 
As a comparison,  the expected theoretical uncertainties
on the inclusive jet cross section from the propagation of pdf uncertainties are below the
10\% level up to a jet $\pt$ of 1 TeV, thus much smaller than the experimental systematics from 
the JES. Therefore it is obvious that a measurement of the inclusive jet cross section
will not allow to constrain the pdfs, unless the JES is known to 2\% or better. This is definitely
beyond reach for the early phase of the LHC, and might remain a huge challenge even later.

Obviously, the knowledge of the JES also has a strong impact on the achievable precision of the
dijet cross section measurement. However, the
problem can be avoided by performing relative instead of absolute cross section measurements. A well suited
observable is the dijet ratio $N(|\eta| < |\eta_\mathrm{in}|) / N(|\eta_\mathrm{in}| < |\eta| < |\eta_\mathrm{out}|)$,
i.e., the ratio of the number of dijet events within an inner region $|\eta| < |\eta_\mathrm{in}|$ to
the number of dijet events within an outer region
$|\eta_\mathrm{in}| < |\eta| < |\eta_\mathrm{out}|$. Both leading jets of the dijet event must satisfy the $|\eta|$ cuts,
with typical values of 
$\eta_\mathrm{in}=0.7$ and $\eta_\mathrm{out}=1.3$. The dijet ratio
has two interesting features. First, it is very sensitive to new physics, such as contact interactions or the
production of a heavy resonance, because those lead to jets at more central rapidities than in genuine QCD 
dijet events. Second, in the ratio we can expect many systematic uncertainties to cancel. For example, the
luminosity uncertainty completely disappears in the ratio. More importantly, also the JES 
uncertainty is strongly reduced, since the dijet ratio is sensitive only to the relative knowledge of the
scale as a function of rapidity, but not to the absolute scale any more. 

As we have seen above, the JES is the dominant source of uncertainty in jet cross section measurements.
Obviously, it is also important for many other analyses and searches which involve jet final states and
possibly invariant mass reconstructions with jets. Therefore major efforts are devoted by the experimental
collaborations to prepare the tools for obtaining JES corrections, both from the Monte Carlo simulations
and, more importantly, from the data themselves. Currently approaches are followed which are inspired
by the Tevatron experience \cite{Bhatti:2005ai, Abbott:1998xw}. The correction procedure is split into several
steps, such as offset corrections (noise, thresholds, pile-up), relative corrections as a function of $\eta$,
absolute corrections within a restricted $\eta$-region, corrections to the parton level, flavour-specific corrections
etc. At the LHC startup we will have to rely on Monte Carlo corrections only, but with the first data coming in
it will be possible to switch to data-driven corrections. At a later stage, after a lot of effort will have
gone into the careful tuning of the Monte Carlo simulations, it might be feasible to use Monte Carlo 
corrections again. For example, a rough estimate for the early JES uncertainty evolution in CMS might be
 10\% at start-up,
7\% after 100 pb$^{-1}$ and 5\% after  1 fb$^{-1}$.  Certainly it will be difficult and
require time to obtain a detailed understanding of the non-Gaussian tails in the jet energy resolution. 

Concerning data-driven JES corrections, one of the best channels is $\gamma +$jet production. At leading
order, the photon and the jet are produced back-to-back, thus the precisely measured photon energy
can be used to balance the jet energy. Real life is more difficult, mainly because of additional 
QCD radiation and the large background from jets faking a photon. These can be suppressed
very strongly  with
tight selection and isolation cuts (e.g., no additional third jet with a transverse energy beyond a certain
threshold and tight requirements on additional charged and neutral energy in a cone around the photon). 
The need to understand well the photon-faking jet background 
and the photon fragmentation is avoided by using the channel Z$(\rightarrow \ell\ell) + $jet, with 
electrons or muons, however, at the price of a lower cross section.

%%%%%%%%%%%%%%%%%%%%%%%%%%%%%%%%%%%%%%
% W/Z production
%%%%%%%%%%%%%%%%%%%%%%%%%%%%%%%%%%%%%%
\subsection{Vector Boson Production}
 \label{sec:dissertori_W/Z} 

The production of vector bosons ($W$ and $Z$), triggered on with their subsequent leptonic
decays, will be among the most important and most precise tests of the SM at the LHC.
The leptonic channels, mainly electrons and muons, can be reconstructed very cleanly, 
at high statistics, with excellent resolution and efficiency 
and very small backgrounds. At the same time, the theoretical
predictions are known to high accuracy, as discussed in more detail below. 
This precision will be useful for constraining pdfs, e.g.\ by measuring the rapidity dependence of the
Z production cross section, in particular when going to large rapidities and thus probing low $x$ values.
As proposed in \cite{Dittmar:1997md}, this process will serve as a standard candle for 
determining to high precision (at the few per-cent level) the proton-proton
luminosity or alternatively the parton-parton luminosity. Finally, it will be attempted to improve
 on the current precision of the $W$ mass. Besides that, $W$ and $Z$ production will be an important experimenter's tool. As mentioned already earlier, $Z$ and $W$ decays to leptons will be used to understand and calibrate various sub-detectors, measure the lepton reconstruction efficiencies and control even the missing transverse energy measurement. 

Below I will first discuss the inclusive case, concentrating on 
resonant production. Then I will highlight some issues for the $W$ and $Z$ production in 
association with jets. Although being highly interesting processes, di-boson production 
will not be discussed here, since for integrated luminosities up to 1 fb$^{-1}$ the statistical
precision will be the limiting factor for these measurements and only allow for a first
proof of existence and rough validations of the model expectations.

%%%%%%%%%%%%%%%%%%%%%%%%%%%%%%%%%%%%%%
% inclusive W and Z
%%%%%%%%%%%%%%%%%%%%%%%%%%%%%%%%%%%%%%

\subsection{Inclusive $W$ and $Z$ production}
 \label{sec:dissertori_inclWZ}

\begin{figure}[htb]
\begin{center}
\includegraphics[width=0.80\textwidth]{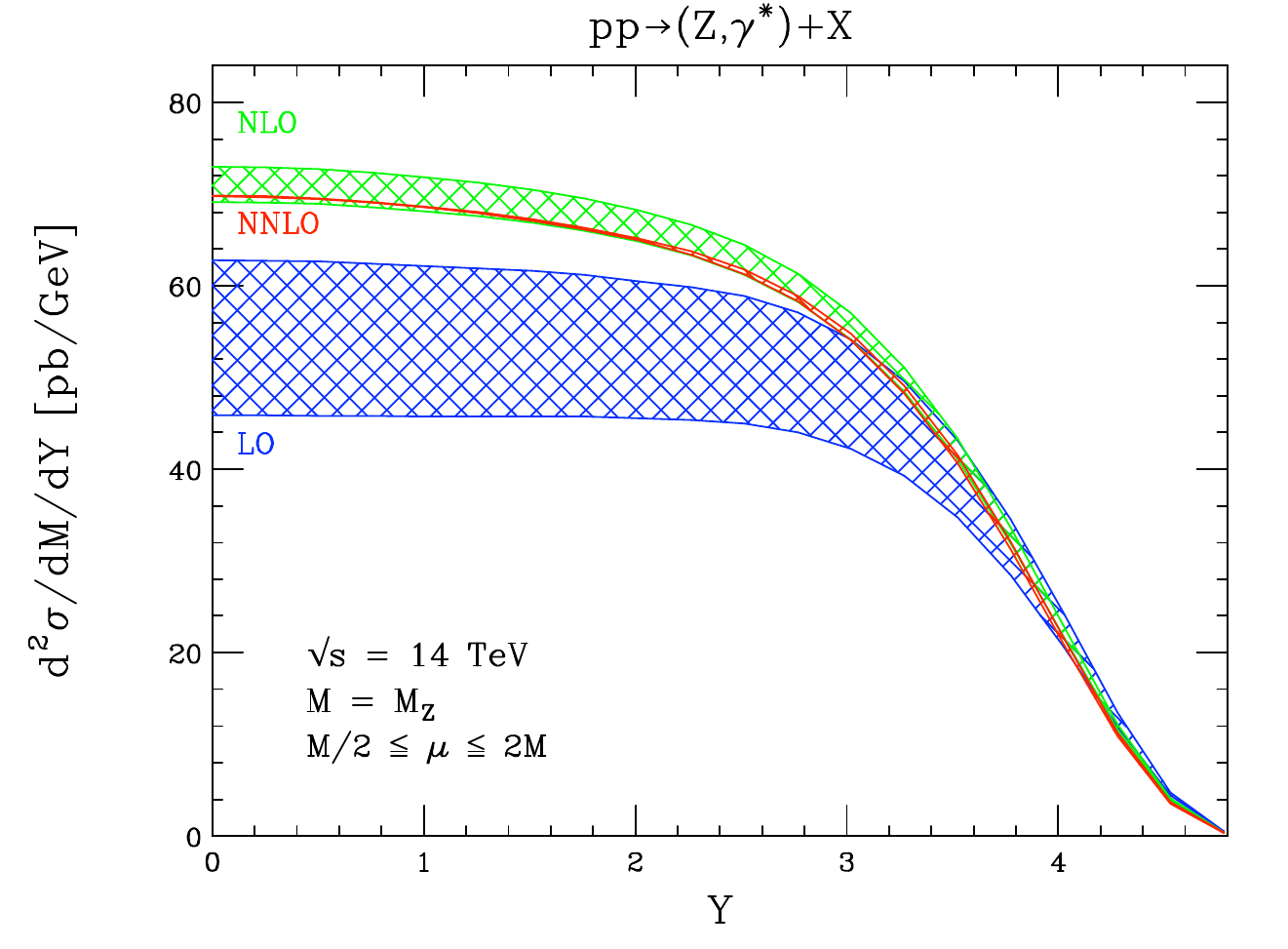}
\caption{QCD predictions at various orders of perturbation theory for the $Z$ rapidity
 distribution at the LHC. The shaded bands indicate the renormalization scale uncertainty.
 Plot taken from \cite{Anastasiou:2003ds}.}
\label{fig:dissertori_zrap}
\end{center}
\end{figure}

Inclusive $W$ and $Z$ production currently is and probably will remain the
theoretically best known process at the LHC. Predictions are available
at next-to-next-to-leading order (NNLO) in perturbative QCD, fully differential in the vector boson and
even the lepton momenta \cite{Melnikov:2006kv}. Figure
\ref{fig:dissertori_zrap} shows the $Z$ rapidity distribution at various
orders in perturbation theory. We see that the shape stabilizes when going
to higher orders and that the NNLO prediction nicely
falls within the uncertainty band of the next-to-leading order (NLO) expansion, giving confidence
in the good convergence of the perturbation series. More importantly, 
the renormalization scale uncertainty is strongly reduced at NNLO, to 
a level of about 1\% for $Z$ rapidities below 3. A renormalization scale uncertainty
even below 1\% can be obtained for ratio observables such as 
$\sigma(W^+)/\sigma(W^-)$ and $\sigma(W)/\sigma(Z)$, possibly
as a function of rapidity. Again, ratio measurements are interesting also from the
experimental point of few, since many systematic uncertainties cancel completely
or to a large extend. The prospect of a precise measurement and 
knowing the hard scattering part of the process so well
means that we have a tool for precisely constraining pdfs (or couplings and masses, in
a more general sense). Indeed, when 
taking the full theoretical prediction for the $W$ and $Z$ production cross section, 
i.e., the convolution of pdfs and hard scattering part,
its uncertainty is dominated by the limited knowledge of the pdfs, estimated
to be below 5\% \cite{Watt:2010qt}. This will then also limit the proton-proton luminosity to
a precision of this size, unless the pdfs are further constrained, mainly by the
rapidity dependence of the cross section, as for example shown in Ref.\ \cite{Dittmar:2005ed}.

In this context one should highlight the importance of having differential
cross section predictions. If we take resonant $W$ and $Z$ production at central
vector boson rapidity, we probe $x$ values of around 0.006, a region rather well
constrained by the current pdf fits. However, for larger rapidities
we probe more and more the small $x$ region, which is less well known, 
e.g., at leading order and for a $Z$ rapidity 
of 3 we need (anti-)quark pdfs at $x=0.12$ and $x=0.0003$. Experimentally,
because of the detector acceptance, we can only access a limited sub-region of the full 
phase space. This means that when measuring a total cross section, we have
to extrapolate the measurement to the full acceptance (e.g., full rapidity), which introduces a 
model dependence, especially on the poorly known low-$x$ region. On the other hand,
having differential predictions, we can compute exactly the same quantity as we
measure, thus eliminating any extrapolation uncertainty. Similarly, for constraining
NLO (NNLO) pdfs, exactly the same acceptance cuts (on the leptons) as in the data can now be applied to
the available NLO (NNLO) predictions. Of course, with more and more differential higher-order 
predictions becoming available, this kind of argument applies to any
cross section measurement (and/or deduced determination of physics quantities such
as couplings, masses, pdfs), namely that we should compare the measurements
and predictions for the experimentally accessible acceptance and avoid unnecessary extrapolations,
which will not teach us anything new and only introduce additional uncertainties.

As mentioned above, the experimental reconstruction of $W$ and $Z$ production is rather straight forward.
Leptons are required to have a minimum $\pt$ of about 20 GeV,
 within a pseudo-rapidity of 2.5. In the $Z$ case the mass peak allows for further event selections and background estimations. However,
the neutrino in the $W$ decay leads to missing energy, which obviously is reconstructed less precisely.
Instead of an invariant mass peak only the transverse $W$ mass can be reconstructed, with larger
backgrounds than for the $Z$.

%%%%%%%%%%%%%%%%%%%%%%%%%%%%%%%%%%%%%%
% W/Z + jets
%%%%%%%%%%%%%%%%%%%%%%%%%%%%%%%%%%%%%%

\subsection{$W/Z$+jets production}
 \label{sec:dissertori_WZjets}

Vector bosons produced in association with jets lead to final states with
high-$\pt$ leptons, jets and possibly missing transverse energy. Such a topology
is also expected for many searches, in particular for squark and gluino
production and subsequent cascade decays. Obviously it will be important to
understand these SM processes as quickly as possible and validate the
available Monte Carlo generators, which typically combine LO matrix elements with parton showers.
A standard observable will be the $W/Z$ cross section as a function of the associated
leading jet transverse momentum or the number of additional jets. Obviously, such measurements
will suffer from the same JES uncertainties as the QCD measurements discussed above, and thus
constitute only limited calibration tools during the early data taking.
The problem can be reduced by defining clever ratios of cross sections, involving different
vector bosons and/or number of additional jets, or by normalizing the predictions to the data
in limited regions of the phase space (e.g.\ for small jet multiplicity and extrapolating to larger multiplicities).
A completely different approach is to take a more inclusive look at this process, in the sense
that the $Z$ transverse momentum is measured from the lepton kinematics, which is possible
at high statistical and, more importantly, high experimental accuracy.
This distribution can be understood as the convolution
of the $Z+0/1/2/\ldots$jets distributions, therefore any model intended to describe $Z$+jets production
has necessarily to reproduce the $Z$ $\pt$ distribution over its full range.

%%%%%%%%%%%%%%%%%%%%%%%%%%%%%%%%%%%%%%
% top production
%%%%%%%%%%%%%%%%%%%%%%%%%%%%%%%%%%%%%%

\section{Top pair production}
 \label{sec:dissertori_top}
 
The top quark is produced very abundantly at the LHC. With 1 fb$^{-1}$ of integrated
luminosity, we should already have a couple of thousand clean signal events on tape
in the di-lepton channel, and a factor of 10 more in the single lepton channel (lepton+jets
channel). The physics case for the study of top production is very 
rich and can not be discussed in detail here. For example, a recent review can be found in Ref.\ \cite{D'hondt:2007aj}. 
Combining many different channels, a top mass measurement with a precision of
1 GeV might be achieved, which together with a precise $W$ mass measurement constitutes an important indirect constraint of SM predictions and its extensions. 
The production cross section (for single and top-pair production) will be an important measurement,
again for testing the SM predictions and because top production is a copious background to a large number
of new physics searches. In the single muon+X channel, the top-pair production cross section
will soon (i.e.\ with about 1 fb$^{-1}$) be measured with a statistical precision of 1\%. The total 
uncertainty of 10-15\% (excluding the luminosity uncertainty) will be dominated by systematics,
most notably due to the knowledge of the b-tagging efficiency. 
Finally, top production will
become an extremely valuable calibration tool. The mass peak can already be
reconstructed with much less than 1 fb$^{-1}$, even without b-tagging requirements.
With a clean sample in hand, it can be exploited for controlling the b-tagging efficiency
and serve as closure test for the JES corrections determined from other processes.
Concerning the JES, the mass of the hadronically decaying $W$ serves as calibration handle.

%%%%%%%%%%%%%%%%%%%%%%%%%%%%%%%%%%%%%%
% Conclusions
%%%%%%%%%%%%%%%%%%%%%%%%%%%%%%%%%%%%%%

\section{Conclusions}
 \label{sec:dissertori_Conclusion} 

It has been an unprecedented challenge to design and construct the LHC experiments,
as well as to put them into operations. Here an attempt was made to sketch the most
important criteria, which were at the basis of the many design choices, as well as to
give a rough comparison of the expected performance of the ATLAS and CMS detectors.
The quality of the data, which resulted from first LHC collisions in late 2009, 
gives strong confidence that excellent physics results should appear soon after
high-energy operations start in 2010. Thus a very exciting period is ahead of us.

%%%%%%%%%%%%%%%%%%%%%%%%%%%%%%%%%%%%%%%%

\section{Acknowledgements}
 
I would like to thank my colleagues in the LHC experiments who helped me in the
preparation of the lectures and these proceedings, in particular D. Treille, 
D. Froidevaux, S. Cittolin, A. Herve, P. Jenni and R. Tenchini.
I would like to thank the organizers of this school for the invitation and their
great hospitality during my stay. It was with great sorrow that I received the
announcement, at the time of writing this summary, that one of the organizers, Th. Binoth,
tragically died in an avalanche accident.

%%%%%%%%%%%%%%%%%%%%%%%%%%%%%%%%%%%%%%%%

%\section*{References}

\end{document}